\documentclass[]{jfm}

\usepackage{graphicx}
\usepackage{caption}
\usepackage{subcaption}
\usepackage{graphicx}
\usepackage{newtxtext}
\usepackage{newtxmath}
\usepackage{natbib}
\usepackage{hyperref}
\hypersetup{
    colorlinks = true,
    urlcolor   = blue,
    citecolor  = black,
}

\newcommand{\RomanNumeralCaps}[1]

\title{Analysis of near wall flame and wall heat flux modeling in turbulent premixed combustion}

\author{Kunlin Li\aff{1}
  ,
  Chenlin Guo\aff{1}
  ,
  Zhaofan Zhu\aff{2}
  ,
  Haiou Wang\aff{2}
 \and Lipo Wang\aff{1}
 \corresp{\email{lipo.wang@sjtu.edu.cn}}
 }

\affiliation{
\aff{1}UM-SJTU Joint Institute, Shanghai Jiao Tong University, Shanghai, China
\aff{2}State Key Laboratory of Clean Energy Utilization, Zhejiang University, Hangzhou, China
}

\begin{document}
\maketitle

\begin{abstract}
Reactive flows in confined spaces involve complex flame-wall interaction (FWI). This work aims to gain more insights into the physics of the premixed near-wall flame and the wall heat flux as an important engineering relevant quantity. Two different flame configurations have been studied, including the normal flushing flame and inclined sweeping flame. By introducing the skin friction vector defined second-order tensor, direct numerical simulation (DNS) results of these two configurations show consistently that larger flame curvatures are associated with small vorticity magnitude under the influence of the vortex pair structure. Correlation of both the flame normal and tangential strain rates with the flame curvature has also been quantified. Alignment of the progress variable gradient with the most compressive eigenvector on the wall is similar to the boundary free behavior. To characterize the flame ordered structure, especially in the near-wall region, a species alignment index is proposed. The big difference in this index for flames in different regions suggests distinct flame structures. Building upon these fundamental insights, a predictive model for wall heat flux is proposed. For the purpose of applicability, realistic turbulent combustion situations need to be taken into account, for instance, flames with finite thickness, complex chemical kinetics, non-negligible near-wall reactions, and variable flame orientation relative to the wall. The model is first tested in an one-dimensional laminar flame and then validated against DNS datasets, justifying the model performance with satisfying agreement.

\end{abstract}


\section{Introduction}
\label{S1}
Energy release from fossil fuels in reactive flows typically occurs in space-confined combustors with complex flame-wall interaction (FWI). Numerous efforts have been made to investigate the fundamental challenges of the underlying physics. Specifically, in the near-wall region advection becomes negligible because of the no-slip velocity condition. As a result, the advective mixing between fuel and oxidizer is significantly suppressed. Thus, flames near the wall are basically premixed~\citep{legier2000dynamically,de2002flame,mohaddes2022hot}. Therefore, it is of particular importance to gain deeper insights into FWI in premixed combustion. 

According to the flow setup, FWI can be differently categorized, such as evolving head-on quenching (HOQ)~\citep{1981Flame, 1981A, 1993Direct, P1997Analysis, Jiawei2016Effects,mohan2025distributions}, side-wall quenching (SWQ)~\citep{clendening1981raman, ezekoye1992increased, alshaalan1998turbulence, bellenoue2003direct, zirwes2021numerical, chen2023study} and the flushing flame quenching (FFQ), where the flame is advected by the incoming flow against the wall at the statistically stationary state~\citep{2018Analysis, ZHAOPRF2019, ZHAO2021PCI, konstantinou2021effects}. In HOQ, the fresh reactants are trapped between the flame and the wall, leading to an unsteady flame propagating towards the wall. In SWQ, the flame propagates nearly parallel to the wall. FWI is important only for the tip part of the flame front. In FFQ, the counterflow-like fresh reactants are fed from the inlet against the solid wall. The flame is anchored at the statistically stationary state with a balanced position under the joint control of the flame speed and inflow strain rate.  

Fundamentally, physical properties of the near-wall flame can be largely different from those of the boundary free flames. For instance, as the flame approaches the cold wall, flame extinction occurs at a critical point where the heat loss exceeds the heat released from the flame zone. An interesting observation is that enhanced turbulence intensity causes the flame front to experience more pronounced distortion and stronger pushing toward the cold wall. Consequently, the near-wall flames are likely fragmented to generate edge flames. As a familiar concept in diffusion combustion~\citep{buckmaster2002edge,karami2016edge,chen2021flame}, a flame edge is defined as the intersection between the mixture fraction isosurface and reactive scalar isosurface, separating the extinguished and burning regions. In premixed or partially premixed combustion, the edge flame behavior has also been addressed, for instance, the influence of ambient temperatures on the edge flame structure in the partially premixed mixing layer~\citep{krisman2015polybrachial} and the edge flame propagation under varying strain rates and Lewis numbers~\citep {clayton2019propagation}. Moreover, \cite{2018Analysis} claimed that in the quenching zone, the flame is largely thickened that locally the flame structure may be more complex by entrainment of small eddies~\citep{2010Turbulent}. Considering the difference in the isosurface displacement speed, a flame zone speed~\citep{2018Analysis} was introduced to describe the movement of the flame as a whole.

Compared with the non-reactive case, FWI leads to much enhanced near-wall temperature gradient and the wall heat flux as well. As a quantity important for both applications and fundamental flame physics, the wall heat loss is closely relevant to flame stability and pollutant emission. The reported wall heat flux, for instance $200\,\mathrm{MW/m^2}$ at the throat of the rocket chamber, can result in extreme thermal stress~\citep{frohlich1993heat} and serious lifetime reduction. Numerous studies have revealed the wall heat flux mechanisms. \cite{2010Turbulent} demonstrated a correlation between wall heat flux and near-wall turbulent vorticity structures. Experimental studies~\citep{kosaka2018wall} suggest that at elevated wall temperatures the quenching distance decreases with increased wall heat flux, attributed to a higher laminar flame speed. Flow topology and wall heat flux statistics were investigated by~\cite{lai2019heat}. \cite{ahmed2023influence} showed that wall heat flux is strongly influenced by flow configurations; the mean friction velocity and wall shear stress vary with both configurations and thermal conditions. \cite{padhiary2023effect} conducted constant-volume chamber experiments investigating pressure and turbulence effects on the wall heat flux peak. Through two-dimensional V-shaped flames, \cite{zhu2024correlation} identified negative correlations between time-averaged wall heat flux and quenching Peclet number, with slope depending on the equivalence ratio. \cite{wang2024three} investigated near-wall quenching and wall heat flux characteristics with the flame geometrical effects in three dimensions.

Numerically, the broad flame and flow structure scales in FWI under the realistic operating conditions cannot be sufficiently resolved in either large eddy simulations (LES) or Reynolds-Averaged Navier-Stokes (RANS) simulations; meanwhile, direct numerical simulations (DNS) are typically intractable. In this sense, reliable prediction models for wall heat flux play an irreplaceable role in applications. To model the wall heat flux in reactive flows, \citet{boust2007thermal,boust2009model} analyzed the energy balance relation in one-dimensional HOQ. At the quenching state, the wall heat flux is equal to the heat flux conducting from the reaction zone to the preheat zone, i.e., the temperature is linear between the flame surface and the wall. In principle, this model works only for the flame quenching case. Inside the turbulent layer, the wall heat flux can also be estimated from the energy conservation relation once the mean reaction rate inside the turbulent boundary layer is reasonably evaluated~\citep{Nilan2022}. In a statistically stationary flushing configuration with single-step chemistry and simplified flow properties, \citet{2018Analysis} proposed to couple the stream function in the upstream region, the flame zone, and the constant vorticity relation in the downstream region along the wall-normal (one-dimensional space) direction. The predicted near-wall temperature field as well as the wall heat flux can satisfactorily match the DNS results. Recently, this model has been extended~\citep{Li2024Evaluation} for both the head-on flame and entrained flame in the context of LES. Over a large variation range of the filter sizes, the modeled wall heat flux is almost filter size independent, showing much-improved predictability, compared with the standard subfilter viscosity results. However, under practical conditions, e.g., complex chemistry and realistic flow properties, the model applicability remains unclear.

The present study focuses on both the fundamental properties of FWI and wall heat flux modeling. In the first topic, we address some new perspectives of the near-wall flame, including the alignment relation among flame fronts defined by different species, interaction between the near-wall flame and velocity, and influence of the wall parallel heat transfer. In the second topic, a wall heat flux model is constructed to adapt to generalized FWI scenarios, for instance, realistic fluid properties, detailed chemical kinetics, and the nonzero lateral heat flux. Finally, we summarize the concluding remarks and the meaningfulness of the present work for practical applications.

\section{Problem formulation and direct numerical simulation (DNS)}

\subsection{Problem formulation}

The continuity, momentum, total energy, and species equations are listed as follows
\begin{equation}
    \frac{\partial{\rho}}{\partial {t}} + {\nabla} \cdot \left( {{\rho} {\boldsymbol{u}} } \right ) = 0,
\label{continuity}
\end{equation}
\begin{equation}
   \frac{\partial{\rho \boldsymbol{u}}}{\partial {t}} +  \nabla\cdot( \rho \boldsymbol{u} \boldsymbol{u}) = \nabla \cdot \boldsymbol{\tau} - \nabla p,
\label{ns}
\end{equation}
\begin{equation}
   \frac{\partial{\rho e_t}}{\partial {t}} +  \nabla\cdot( \rho \boldsymbol{u} e_t) = -\nabla \cdot (\boldsymbol{u} p) - \nabla \cdot \boldsymbol{q} + \nabla \cdot(\boldsymbol{\tau}\cdot u) ,
\label{et}
\end{equation}
\begin{equation}
   \frac{\partial{\rho Y_{k}}}{\partial {t}} +  \nabla\cdot( \rho \boldsymbol{u} Y_k) = -\nabla \cdot \boldsymbol{J}_{k} + \Dot{\omega}_k ,
\label{Yi}
\end{equation}
where $\rho$ is the flow density, $\boldsymbol{u}$ is the velocity vector, $Y_k$ is the mass fraction of the $k$-th species and $e_t$ is the total energy. On the right hand side of the above equations, $\boldsymbol{\tau}$ is the stress tensor, $p$ is pressure, $\boldsymbol{q}$ indicates thermal diffusive flux and $\boldsymbol{J}_{k}$ is the species diffusive flux, together with the reaction rate $\Dot{\omega}_k$.

To facilitate the analysis, we first introduce the non-dimensional quantities $(\tilde \cdot)$ with reference scales. The reference length is the domain size $L$ in the wall-normal direction, the reference velocity is the laminar flame speed $S_L^0$, and the reference time is $L/S_L^0$. Progress variable $c$ is defined upon the mass fraction of a reactant $Y_R$ as $(Y_{R,u}-Y_R)/(Y_{R,u}-Y_{R,b})$, where subscripts $u$ and $b$ denote the unburnt state and burnt state, respectively. Temperature $T$ is non-dimensionalized as $\tilde{T}=(T-T_u)/(T_{ad}-T_u)$ with $T_{ad}$ representing the adiabatic flame temperature of the given flow. The wall heat flux $Q_w$ is non-dimensionalized by the flame power $\rho_u C_{p,u} S_L^0 \left( T_{ad} - T_{u}\right)$.  The heat release rate $\Dot \omega_T$ and the thermal diffusive flux $\boldsymbol{q}$ are non-dimensionalized by $\rho_u C_{p,u} S_L^0 \left( T_{ad} - T_{u}\right)/L$. All the involved thermal-physical properties, i.e., heat capacity $C_p$, density $\rho$, thermal conductivity $\lambda$, and diffusivity for species $D_k$, use reference values of the corresponding unburnt gas. For convenience, the tilde sign for non-dimensional quantities is omitted hereafter without explicit mention. For comparison, two different premixed turbulent combustion setups are studied in the following.

\subsection{Wall normal flushing flame}
\begin{figure}
    \centering
    \subfloat[]{\includegraphics[width=0.48\textwidth]{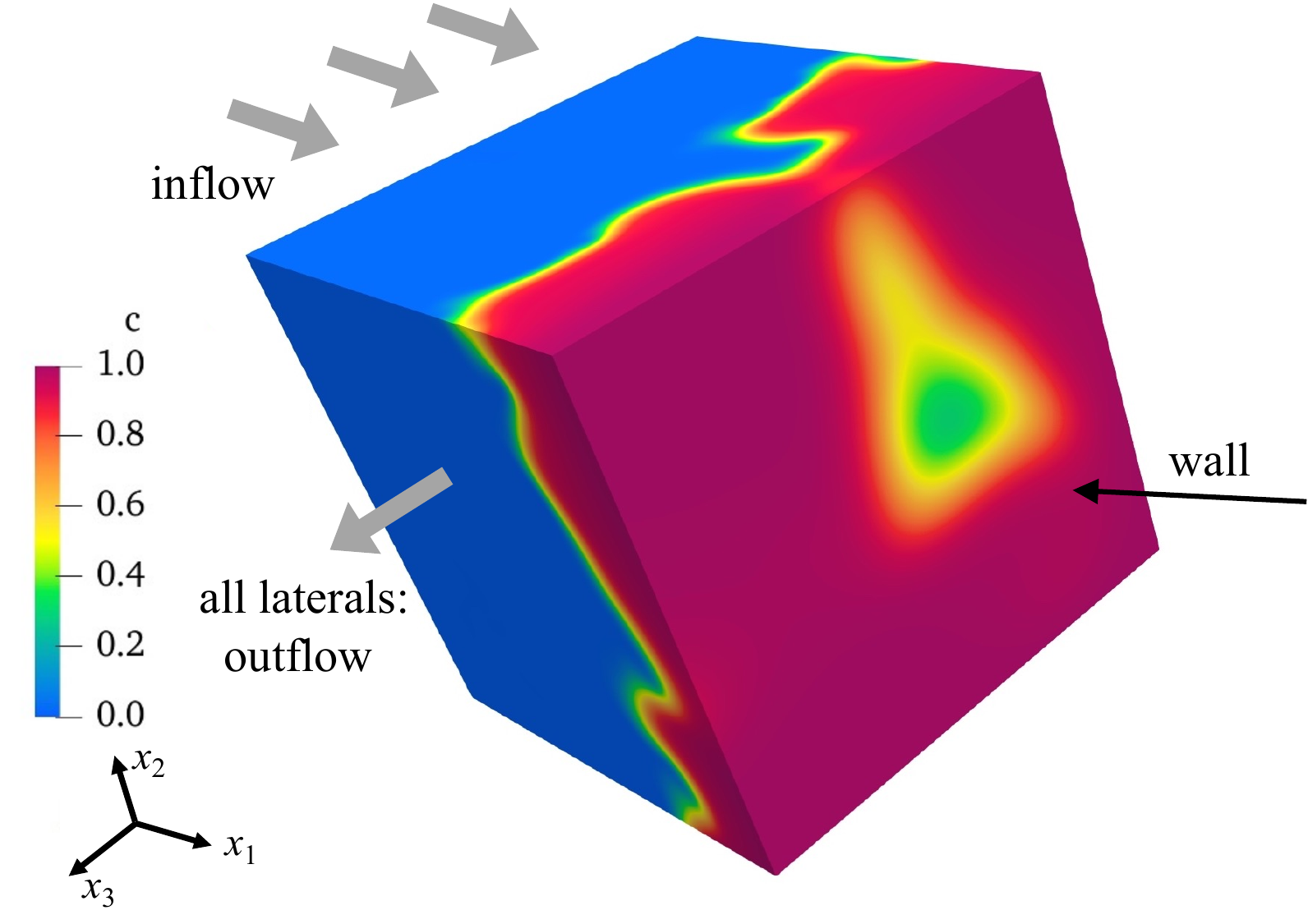}\label{fig:DXDNSC}}
    \subfloat[]{\includegraphics[width=0.45\textwidth]{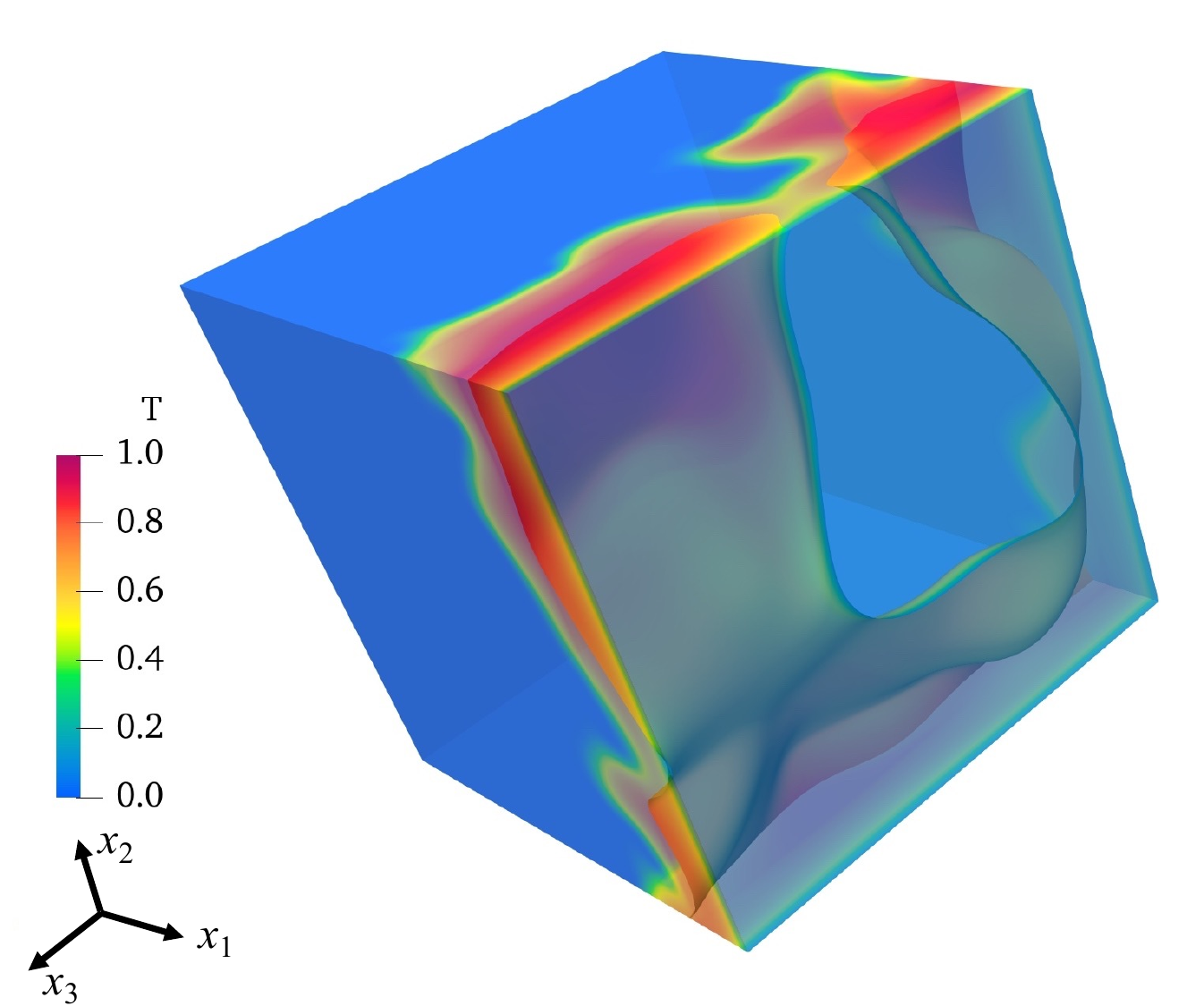}\label{fig:DXDNST}}
    \caption{Spatial distribution of: (a) instantaneous hydrogen-based progress variable field; (b) instantaneous temperature field together with the flame isosurface defined by $c=0.8$.}
    \label{fig:DXDNS}
\end{figure}

The first one is wall normal flushing flame. As shown in Fig.~\ref{fig:DXDNS}, the fresh reactant as inflow is fed into the domain against the wall normal direction $x_1$ and flows out through the lateral boundaries along $x_2$ and $x_3$. At the statistically stationary state, the mean flame-wall distance is strain rate dependent. In a series of recently finished studies, both the simplified single-step reaction mechanism with constant thermal-physical properties~\citep{2018Analysis} and multi-step kinetics with realistic thermal-physical properties~\citep{zhao2022near} have been investigated. It is found that the multi-step chemistry leads to more complex flame characteristics, including local quenching, flame thickening, and various flame topologies. 

In the present work, a premixed hydrogen-air reactive flow is solved at atmospheric pressure using the ‘KARFS’ solver~\citep{perez2018direct,desai2021direct}. The fresh premixed reactant with stoichiometric ratio $\phi=0.7$ and temperature of $300\,\mathrm{K}$ is injected along the $x_1$ direction. The mean inlet velocity is $12\,\mathrm{m/s}$ and the fluctuating magnitude is $6.75\,\mathrm{m/s}$, generated by scanning a prescribed isotropic homogeneous turbulence field~\citep{rogallo1981numerical}. The wall boundary is isothermal and inert with a temperature of $450\,\mathrm{K}$. A 23-step reaction mechanism of the hydrogen-air combustion~\citep{burke2012comprehensive} and the mixture-average transport properties are adopted. In a cubic domain with $3.84\,\mathrm{mm}$ long on each side, equally spaced $256^3$ grids ensure to resolve both the flame and the turbulent scales. 

The flame front is identified as the hydrogen-based progress variable $c=0.8$ isosurface, corresponding to the peak heat release rate under laminar conditions. In addition, extinct regions on $c=0.8$ are excluded if the local $\text{OH}$ mass fraction on the flame front falls below $10\%$ of the maximum $\text{OH}$ mass fraction~\citep{wang2021turbulence}. The reference burnt and unburnt hydrogen mass fractions for calculating the progress variable are $0$ and $0.0204$, respectively. The flame front can be defined based on other species, for instance, the oxygen-based progress variable isosurface $c_{\mathrm{O}_2}=0.64$, where the burnt and unburnt oxygen mass fractions are $0$ and $0.232$, respectively. The distance from the identified position to the wall is defined as the flame wall distance $\delta_f$. Fig.~\ref{fig:DXDNS} (a) and (b) present the overall flow structure, including the hydrogen-based progress variable and the temperature field together with the flame front. Overall, the spatial distribution of the progress variable is similar to that of the temperature, because both quantities follow similar governing equations. However, a clear distinction appears in the vicinity of the cold wall because of their different wall boundary conditions and the differential diffusion effect. To ensure statistical convergence, we collect $15$ snapshots over $2$ to $3$ through-pass times at the statistically stationary state for data analyses.

\subsection{Inclined sweeping flame }
\begin{figure}
    \centering
    \includegraphics[width=0.7\textwidth]{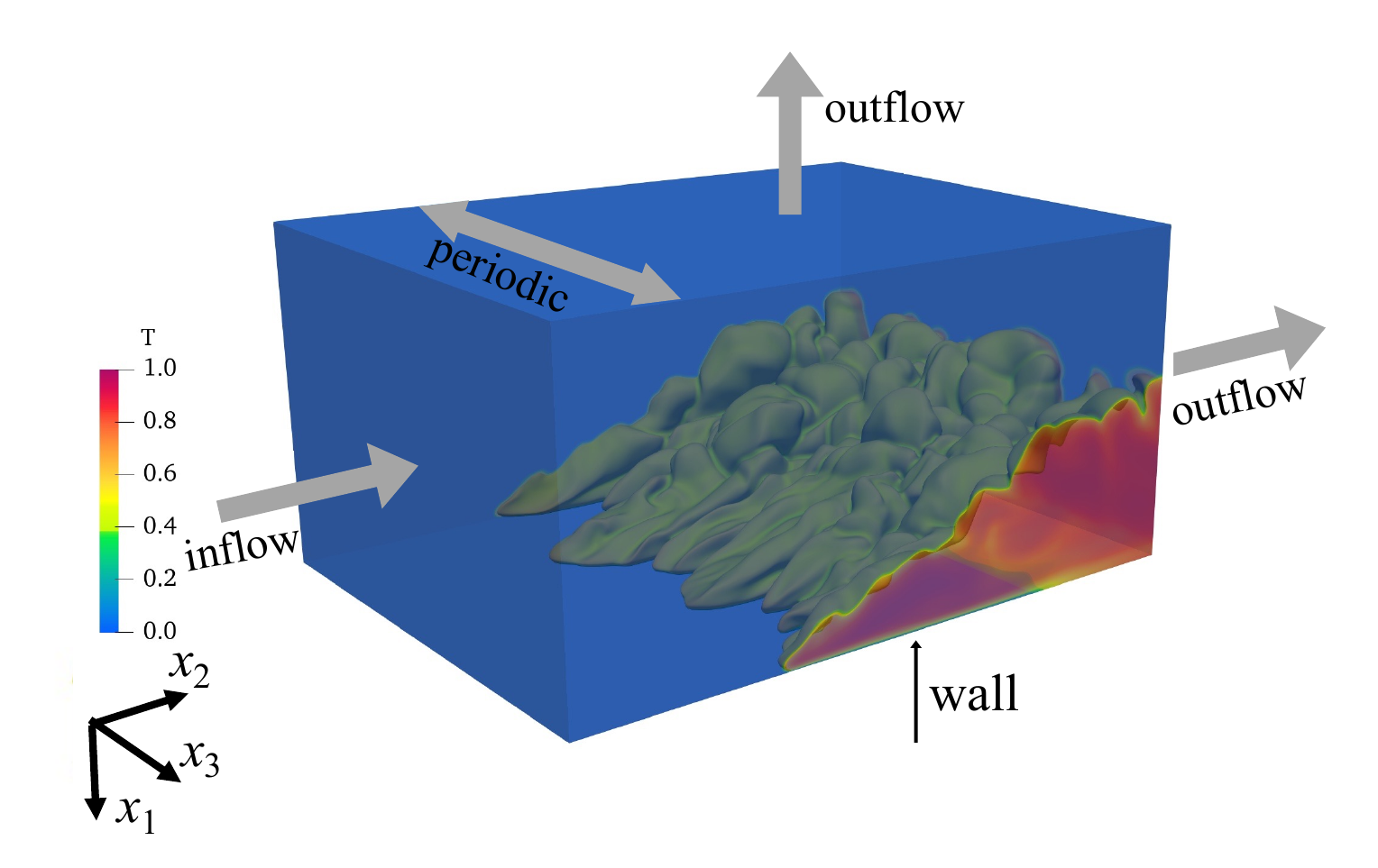}
    \caption{Spatial distribution of an instantaneous DNS temperature field superimposed with a hydrogen-based progress variable $c=0.8$ isosurface.}
    \label{fig:SWQdemo}
\end{figure}

Another setup is the premixed flame flashback over a horizontal flat plate~\citep{chen2023study}. As shown in Fig.~\ref{fig:SWQdemo}, in a box domain with the wall normal along $x_1$, the premixed reactant comes into the domain from the left inlet along $x_2$. After ignition, the flame travels back and interacts with the turbulent boundary layer. As can be seen in Fig.~\ref{fig:SWQdemo}, the flame zone expands along the main stream, forming an overall inclination angle.

DNS parameters are listed as follows. At the left inlet, a lean premixed hydrogen-air mixture with an equivalence ratio $\phi=0.8$ and inlet temperature $500 \, \mathrm{K}$ sweeps over a flat plate, which is isothermal at $500 \, \mathrm{K}$ in the front part and adiabatic in the rear part to stabilize the flame. The pressure remains at $2 \, \mathrm{atm}$. The Lewis number of each species is constant, calculated from the mixture-averaged transport model of a premixed laminar flame. A 19-step mechanism~\citep{li2004updated} is adopted for combustion chemistry. The numerical solver is ‘S3D’~\citep{chen2009terascale} with a fourth-order Runge-Kutta scheme for time advancing~\citep{kennedy1994several} and an eighth-order finite difference scheme for spatial discretization. The domain size along $(x_1,x_2,x_3)$ is $(L_{x1} =10\,\mathrm{mm}, L_{x2}=20\,\mathrm{mm}, L_{x3} = 15\,\mathrm{mm})$ with grid number $(N_{x1}=480, N_{x2}=1000, N_{x3} =750)$. Specifically, grids in $x_2$ and $x_3$ directions are uniform, while the local refinement in the wall-normal direction $x_1$ ensures to resolve the near-wall flame and boundary layer structure. The boundary condition is periodic along the $x_3$ direction and non-reflecting for other outflow surfaces. The inlet velocity is carefully fabricated by using the real-time velocity at a given streamwise slice of a running non-reacting turbulent boundary layer DNS flow, whose free stream velocity is $40 \, \mathrm{m/s}$. Consistent with the wall normal flushing flame case, the flame front is defined here by the same $c$ isosurface. The reference burnt and unburnt hydrogen mass fractions are $0$ and $0.023$, respectively, while the reference burnt and unburnt oxygen mass fractions are $0$ and $0.227$, respectively.

\section{Analyses of near-wall flame physics}
\subsection{Laminar wall stagnation flame}
\label{s3p1}
First, the quasi one-dimensional laminar wall stagnation flame is studied as a reference. Similar to the wall normal flushing flame setup, a stagnation premixed hydrogen-air flame is injected from the inlet along $x_1$. Under atmospheric pressure the equivalence ratio is $\phi=0.7$, the inlet gas temperature is $300\,\mathrm{K}$, and the wall temperature is constant $450\,\mathrm{K}$. At different inlet strain rates $\varepsilon=u_{\text{in}}/L$, where $u_{\text{in}}$ is the inlet velocity, the laminar flame is anchored at different wall distances. The chemical kinetics includes $9$ species and $23$ elementary reactions~\citep{burke2012comprehensive}.

\begin{figure}
     \begin{subfigure}[]{0.45\textwidth}
         \centering
         \includegraphics[width=\textwidth]{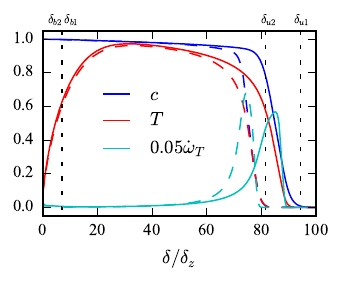}\vskip -8pt
         \label{fig:basic2}
         \caption{$\varepsilon = 1.55$}
     \end{subfigure}
     \begin{subfigure}[]{0.45\textwidth}
         \centering
         \includegraphics[width=\textwidth]{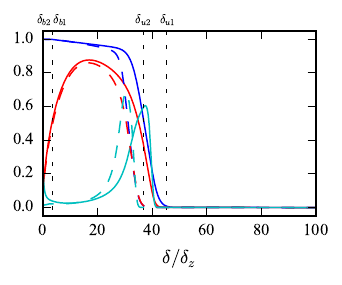}\vskip -8pt
         \label{fig:basic6}
         \caption{$\varepsilon = 4.50$}
     \end{subfigure}
    
     \begin{subfigure}[]{0.45\textwidth}
         \centering
         \includegraphics[width=\textwidth]{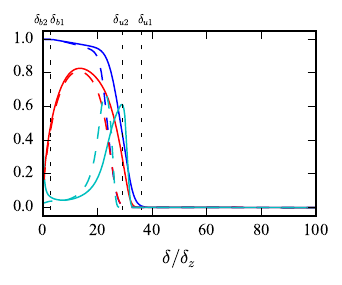}\vskip -8pt
         \label{fig:basic8}
         \caption{$\varepsilon = 5.97$}
     \end{subfigure}
     \begin{subfigure}[]{0.45\textwidth}
         \centering
         \includegraphics[width=\textwidth]{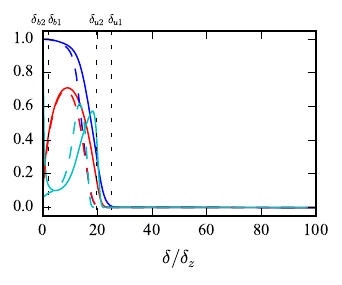}\vskip -8pt
         \label{fig:basic12}
         \caption{$\varepsilon = 8.92$}
     \end{subfigure}
\caption{Distributions of the flame temperature, progress variable (represented by the hydrogen mass fraction), and the heat release rate distributions with respect to the non-dimensional wall distance $\delta/\delta_z$ under different non-dimensional inlet strain rates $\varepsilon$. Results from the mixture-averaged transport model and the unity Lewis number model are shown by the solid and dashed lines, respectively.}
\label{fig:Basic}
\end{figure}

Fig.~\ref{fig:Basic} shows the flame structure varying with $\varepsilon$, where the progress variable $c$ is set based on the hydrogen mass fraction. The flame zone is the reaction part between the left and right edges, where the progress variable gradient is negligibly small, e.g., $\nabla c = 0.01\nabla c_{max}$. For comparison, results from both the mixture-averaged species transport properties (in solid lines with subscript 1) and the unity Lewis number model (in dashed lines with subscript 2) are presented. The spatial point wall distance $\delta$ is normalized by Zel'dovich flame thickness $\delta_z=D_{th}/S_L^0$, in which $D_{th}$ is the unburnt gas thermal diffusivity. $\delta_u$ and $\delta_b$ denote the wall distance from the flame zone edge at the unburnt side and burnt side, respectively. As the strain rate increases, the flame approaches the wall boundary at $\delta=0$; meanwhile, the maximum temperature decreases. For the present chemical mechanism, the reaction zone is clearly thicker than $\delta_z$ ($\sim 10\delta_z$). A sharp difference between the two result sets is that the near-wall heat release rate $\Dot \omega_T$ from the unity Lewis number transport model decreases gradually, but that from the mixture-averaged transport model reaches a local peak, especially when the inlet strain rate is larger. Due to the high diffusivity of $\text{H}_2$, the flame zone from the mixture-averaged model is larger, while $\delta_{{b1}}$ and $\delta_{{b2}}$ are almost identical.

\begin{figure}
\centering\includegraphics[width=0.5\textwidth]{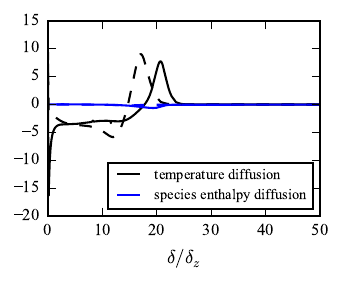}
\caption{Comparison of the non-dimensional diffusion terms in the temperature equation at $\varepsilon=8.92$. The solid and dashed lines indicate the results from the mixture-averaged transport model and the unity Lewis number model, respectively.}
    \label{fig:diffusionCom}
\end{figure}

From the temperature equation along $x_1$
\begin{equation}
   \rho u_1 C_p\frac{\partial T}{\partial x_1} = \frac{1}{RePr}\frac{\partial}{\partial x_1}\left(\lambda \frac{\partial T}{\partial x_1}\right) -\frac{1}{ReSc}\sum_k J_k\frac{\partial h_k}{\partial x_1} + \Dot{\omega}_T ,
\label{TOri}
\end{equation}
the relative contributions of the temperature diffusion term (the first term on the right-hand side) and the species enthalpy diffusion term (the second term on the right-hand side) can be compared. Here $h_k$ is the enthalpy of $k$-th species, with a reference value of $C_{p,u}(T_{ad}-T_{u})$. $Re$, $Pr$, and $Sc$ are the Reynolds number, Prandtl number, and Schmidt number, respectively. Fig.~\ref{fig:diffusionCom} demonstrates that regardless of the transport model selected, the species enthalpy diffusion term remains negligible relative to the temperature diffusion term.

\subsection{Flame and velocity interaction: from space to wall boundary}
The spatial flame and turbulence interaction has been extensively studied using the strain rate tensor. Basically, the most compressive eigenvector dominates species diffusion along the flame normal, while the extensive eigenvector is relevant to the flame corrugation. 

With respect to the isothermal wall, the skin friction vector $\mathbf{F}$ is defined as
\begin{equation}
    \mathbf{F} = [\frac{\partial u_{2}}{\partial N},\frac{\partial u_{3}}{\partial N}],
\end{equation}
where $u_{2},u_{3}$ are the wall tangential velocity components, and $N$ is the wall normal direction (pointing towards the fluid). Moving from space to the wall boundary, the velocity gradient tensor then degenerates to a second-order tensor $\{\mathbf{F}_{ij}\}$, where $\mathbf{F}_{ij}=\partial {\mathbf{F}}_i/\partial x_j\ (i,j\in\{2,3\})$. One can decompose $\{\mathbf{F}_{ij}\}$ as
\begin{equation}
    \{\mathbf{F}_{ij}\} = \{\mathbf{S}_{ij}\} + \{\mathbf{A}_{ij}\},
\end{equation}
where ${\mathbf{S}_{ij}}$ and ${\mathbf{A}_{ij}}$ represent the symmetric part and antisymmetric part, respectively. Specifically,
\begin{equation}
    \mathbf{S}_{ij} = \frac{1}{2}(\frac{\partial \mathbf{F}_i}{\partial x_j} + \frac{\partial \mathbf{F}_j}{\partial x_i}),
\end{equation}
\begin{equation}
    \mathbf{A}_{ij} = \frac{1}{2}(\frac{\partial \mathbf{F}_i}{\partial x_j} - \frac{\partial \mathbf{F}_j}{\partial x_i}).
\end{equation}
$\frac{\partial F_3}{\partial x_2} - \frac{\partial F_2}{\partial x_3}$, denoted as $\omega_1$, means the vorticity normal to the wall. The reference value for $\mathbf{S}_{ij},\ \mathbf{A}_{ij}$ and $\omega_1$ is $S_l^0/L^2$.

\begin{figure}
\centering
     \begin{subfigure}[]{0.49\textwidth}
         \centering
         \includegraphics[width=\textwidth]{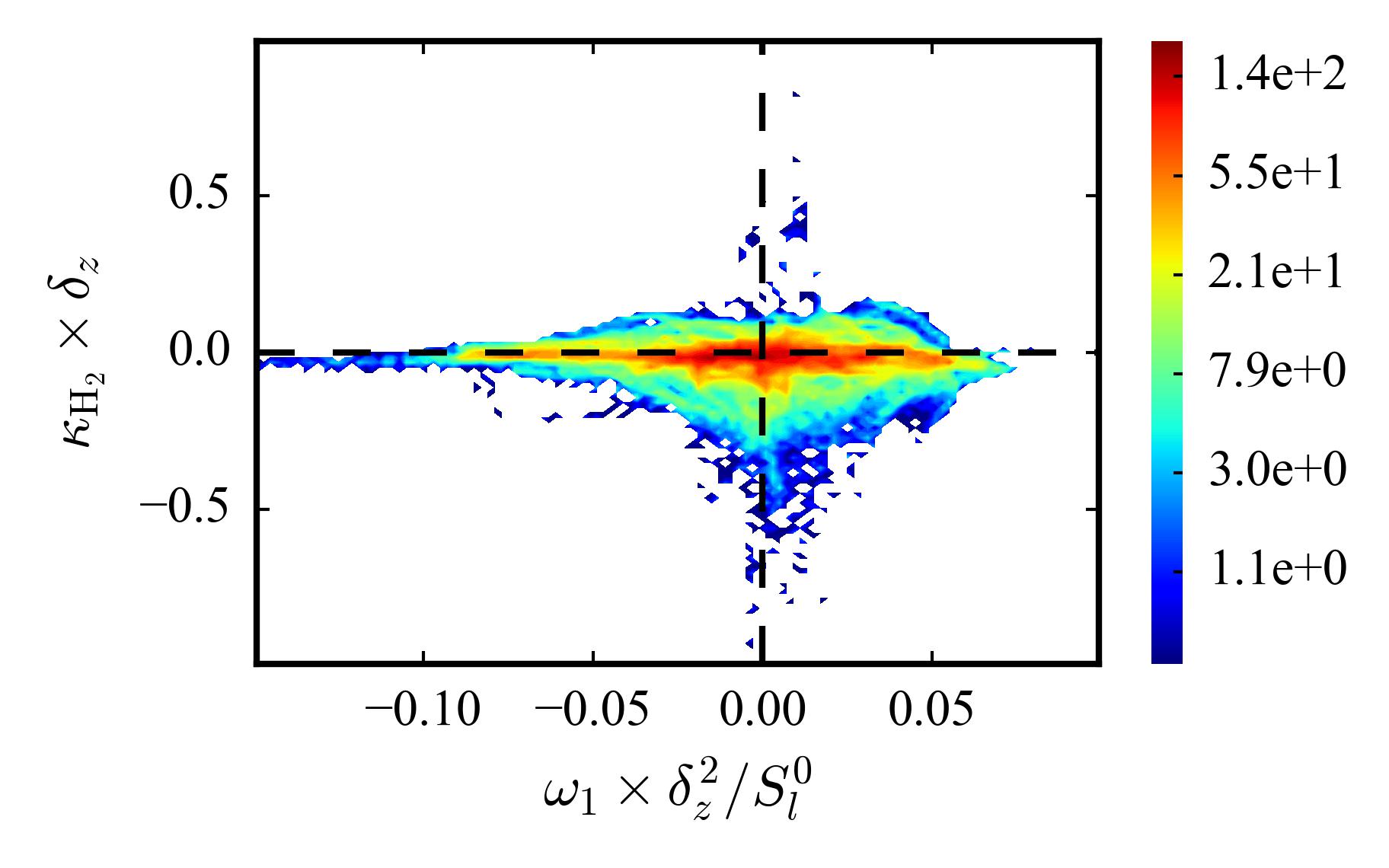}\vskip -8pt
         \caption{}
     \end{subfigure}
     \begin{subfigure}[]{0.49\textwidth}
         \centering
         \includegraphics[width=\textwidth]{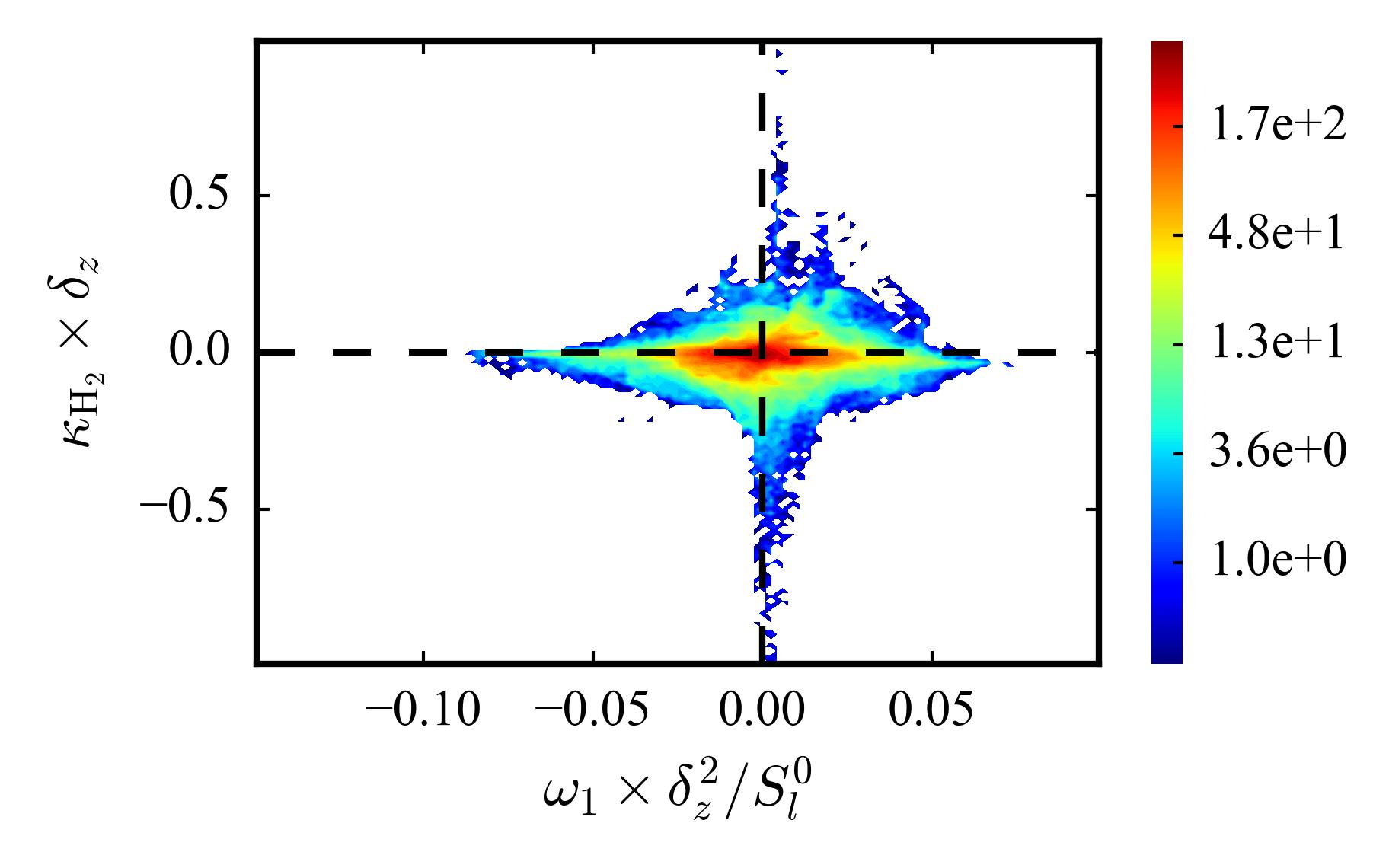}\vskip -8pt
         \caption{}
     \end{subfigure}
     \begin{subfigure}[]{0.49\textwidth}
         \centering
         \includegraphics[width=\textwidth]{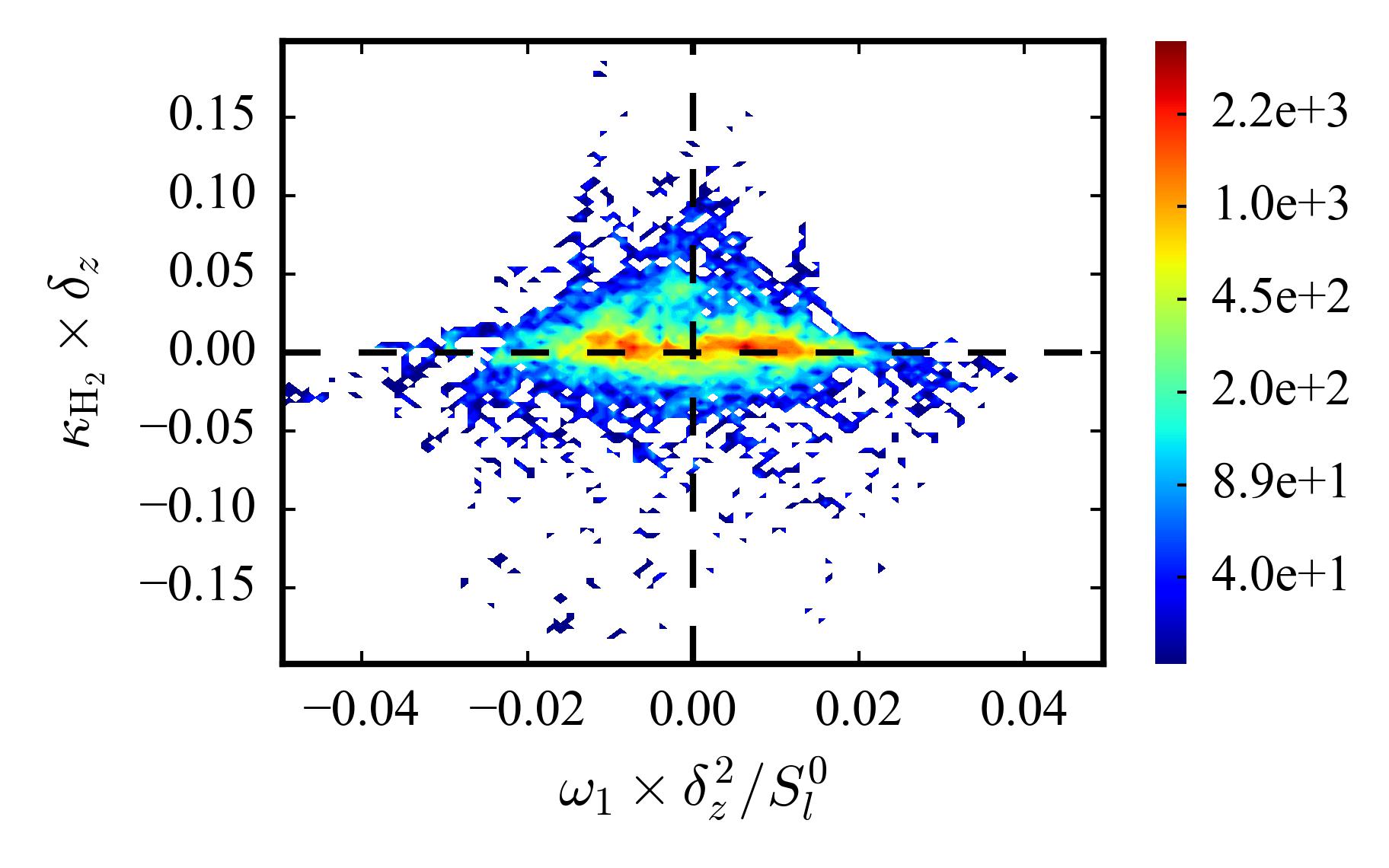}\vskip -8pt
         \caption{}
     \end{subfigure}
     \begin{subfigure}[]{0.49\textwidth}
         \centering
         \includegraphics[width=\textwidth]{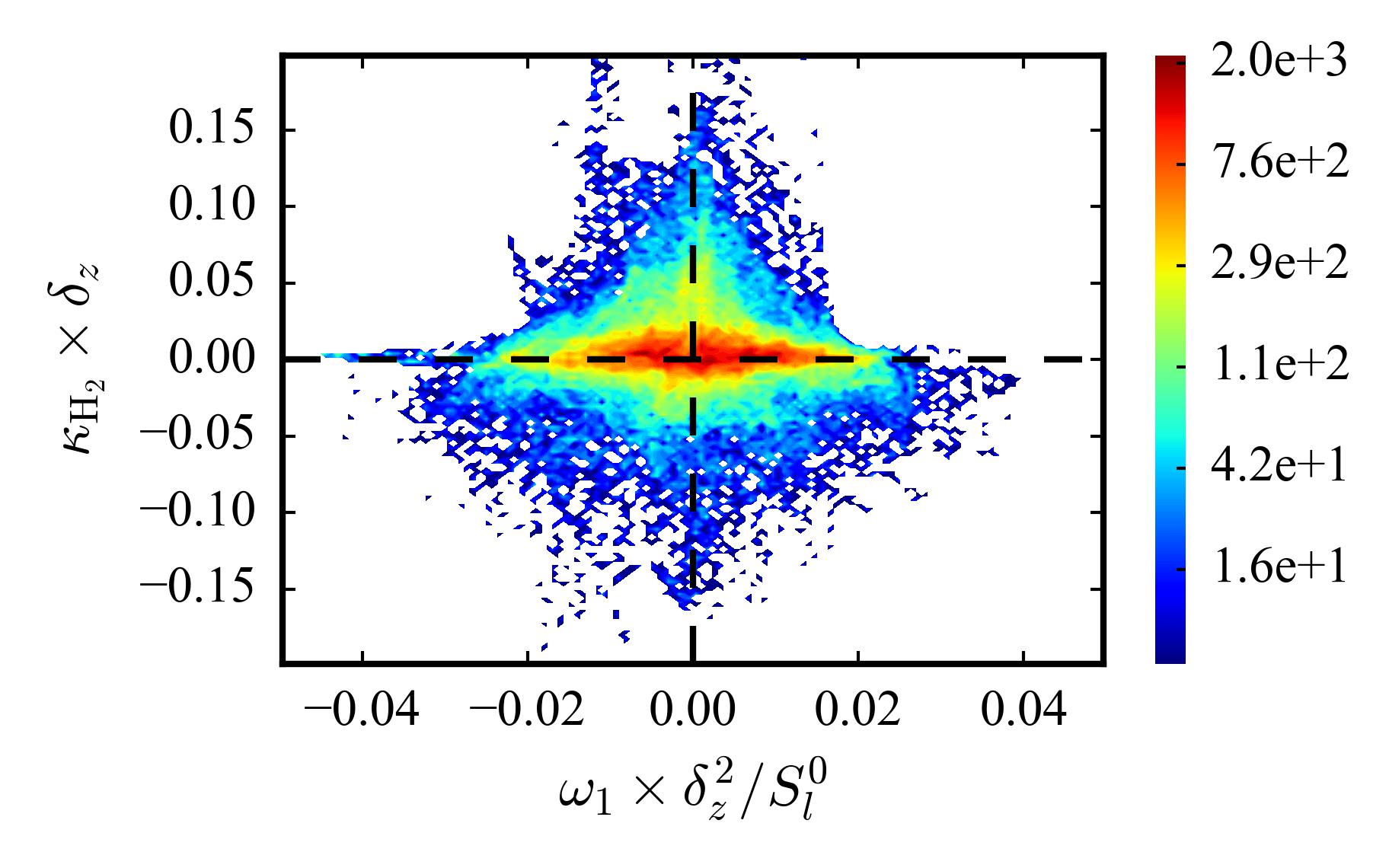}\vskip -8pt
         \caption{}
     \end{subfigure}
\caption{Joint PDFs between $\mathrm{H}_2$ contour line defined curvature $\kappa_{\mathrm{H}_2}$ and $\omega_1$ for the normal flushing flame: (a) $0.65< c\leq0.8$; (b) $0.8 <c\leq0.95$, and for the inclined sweeping flame: (c) $0.65 <c\leq0.8$; (d) $0.8< c\leq0.95$.}
\label{fig:CurOmegaH2}
\end{figure}

\begin{figure}
\centering
     \begin{subfigure}[]{0.49\textwidth}
         \centering
         \includegraphics[width=\textwidth]{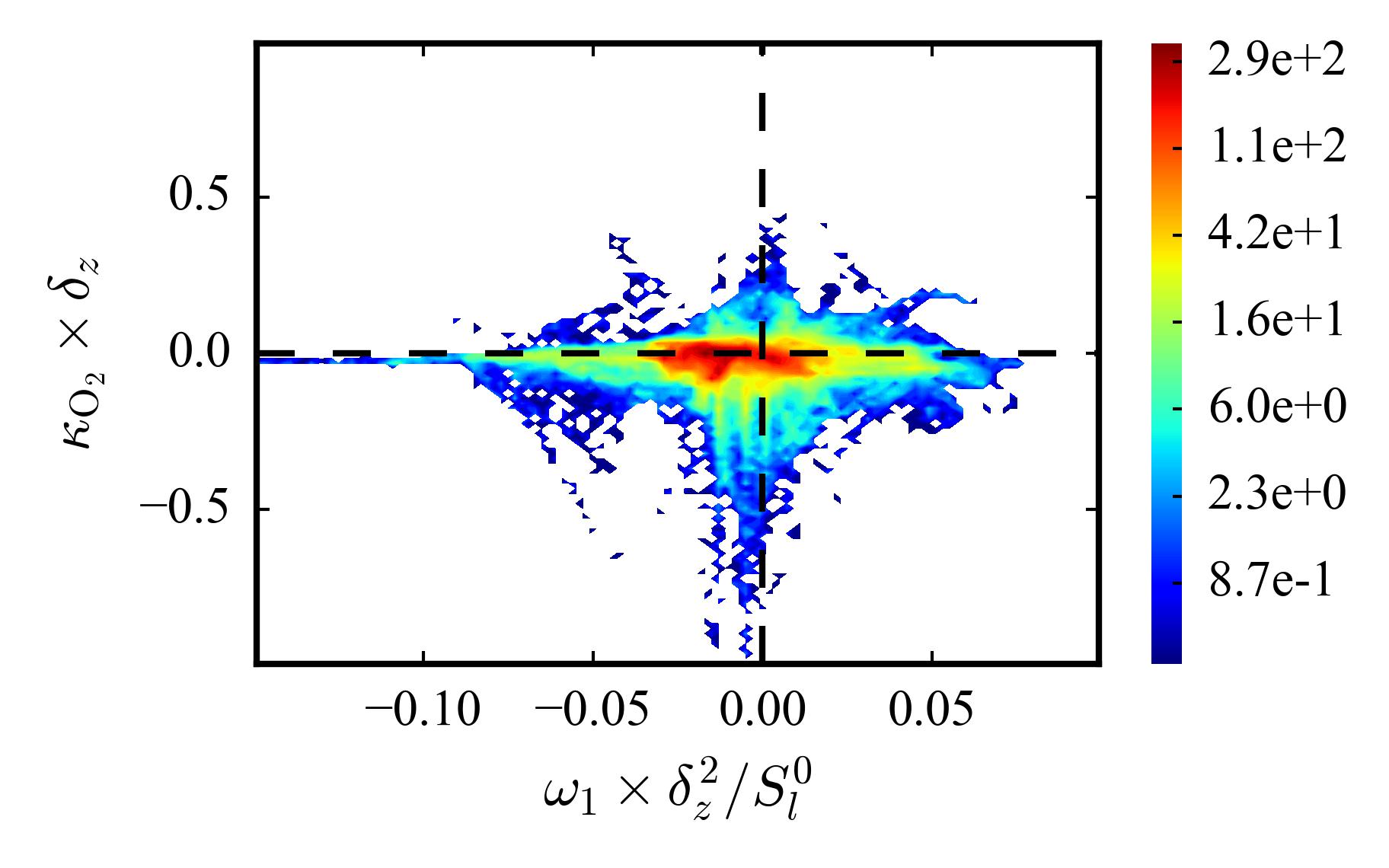}\vskip -8pt
         \caption{}
     \end{subfigure}
     \begin{subfigure}[]{0.49\textwidth}
         \centering
         \includegraphics[width=\textwidth]{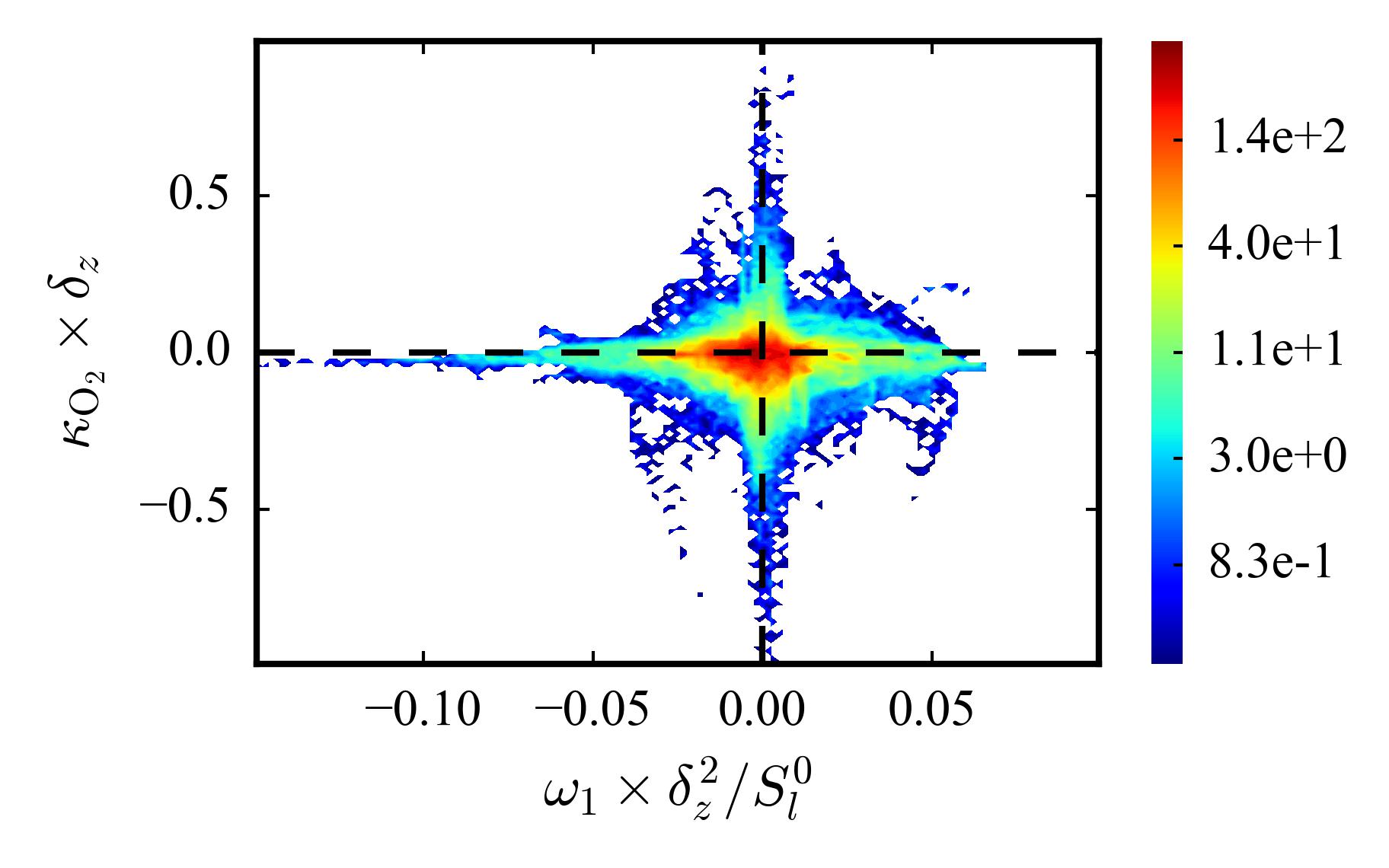}\vskip -8pt
         \caption{}
     \end{subfigure}
     \begin{subfigure}[]{0.49\textwidth}
         \centering
         \includegraphics[width=\textwidth]{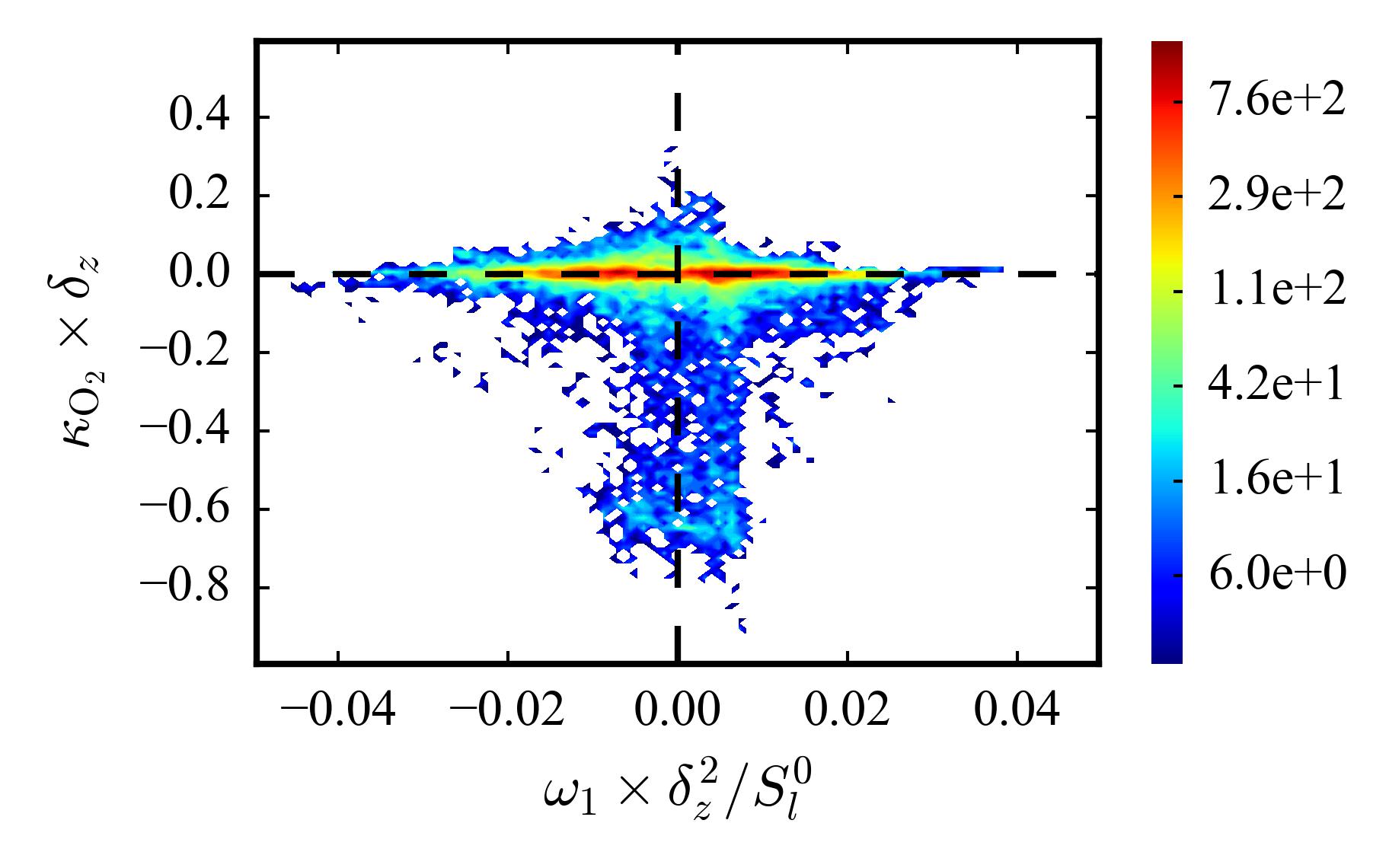}\vskip -8pt
         \caption{}
     \end{subfigure}
     \begin{subfigure}[]{0.49\textwidth}
         \centering
         \includegraphics[width=\textwidth]{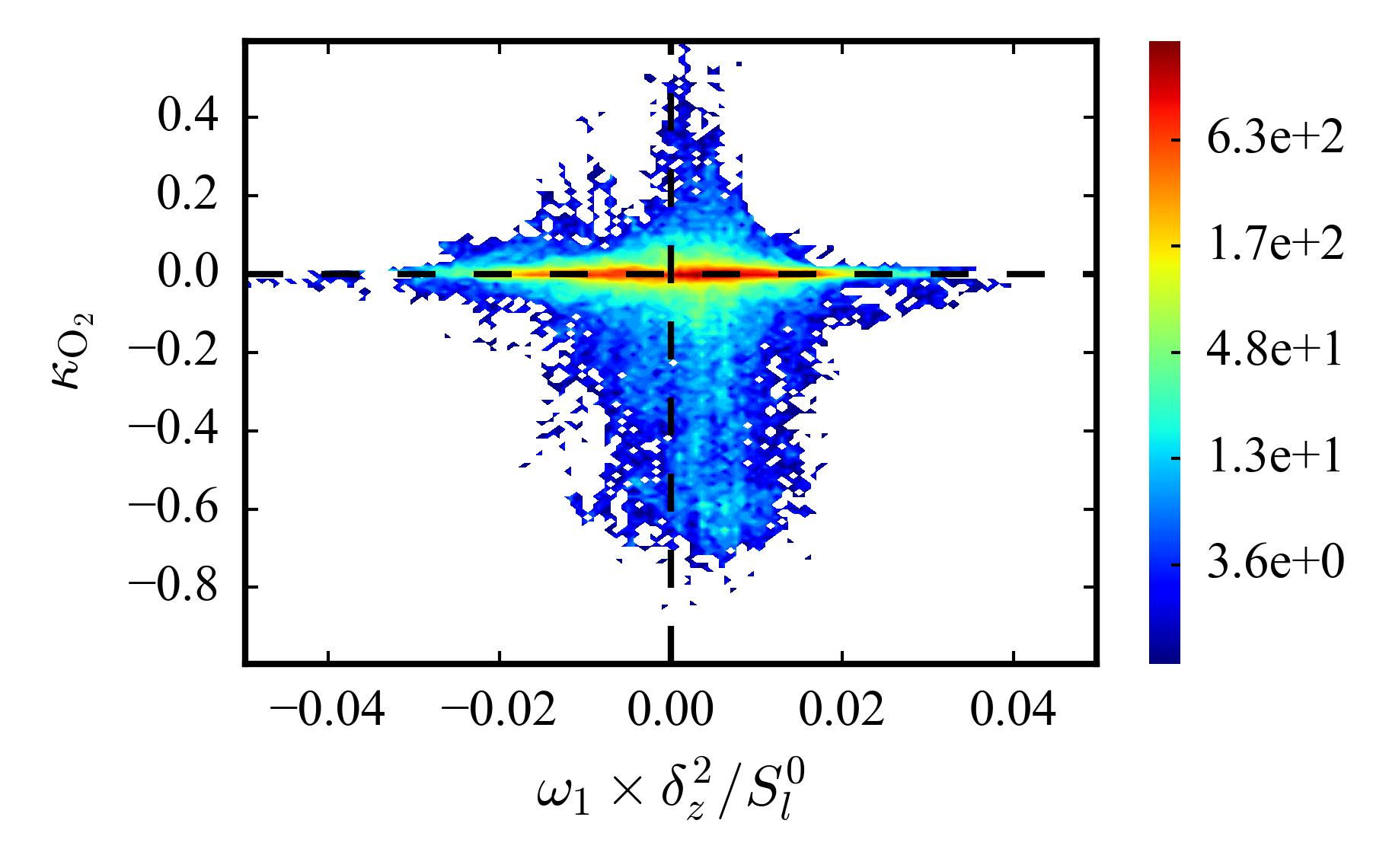}\vskip -8pt
         \caption{}
     \end{subfigure}
\caption{Joint PDFs between $O_2$ contour line defined curvature $\kappa_{\mathrm{O}_2}$ and $\omega_1$ for the normal flushing flame: (a) $0.49< c_{\mathrm{O}_2}\leq0.64$; (b) $0.64 <c_{\mathrm{O}_2}\leq0.79$; and for inclined sweeping flame: (c) $0.49 <c_{\mathrm{O}_2}\leq0.64$; (d) $0.64< c_{\mathrm{O}_2}\leq0.79$.}
\label{fig:CurOmegaO2}
\end{figure}

Such a second-order tensor is important to understand the near-wall flame and velocity interaction. The flame curvature is quantified as
$\kappa=\nabla\cdot\nabla\boldsymbol{n}$, where $\boldsymbol{n}$ is the flame normal pointing toward the unburnt side. By definition, the curvature is positive when the flame is convex towards the unburnt side. The joint PDFs between $\omega_1$ and the flame curvature represented by the $\mathrm{H}_2$ contour lines are shown in Fig.~\ref{fig:CurOmegaH2}. For comparison, results for the $\mathrm{O}_2$ contour lines are also shown in Fig.~\ref{fig:CurOmegaO2}. Since on the wall boundary small progress variable is relatively rare, here samples are collected in relatively broader ranges, including $0.65< c\leq0.8$ and $0.8 <c\leq0.95$. It can be seen that all the joint PDFs peak at the origin, where both the curvature magnitude and vorticity magnitude are low. Large curvature magnitude is inclined to have a small vorticity magnitude. Across the flame, i.e., from Fig.~\ref{fig:CurOmegaH2} (a) to Fig.~\ref{fig:CurOmegaH2} (b) and from Fig.~\ref{fig:CurOmegaH2} (c) to Fig.~\ref{fig:CurOmegaH2} (d), $\omega_1$ is more concentrated in a narrows range, because the flame weakens the vorticity, both in space and on the wall. The similar statistical results for oxygen are shown in Fig.~\ref{fig:CurOmegaO2}, where $0.49< c_{\mathrm{O}_2}\leq0.64$ and $0.64 <c_{\mathrm{O}_2}\leq0.79$ represent two different sides of the flame front. The overall features are similar to Fig.~\ref{fig:CurOmegaH2}, except that the large negative curvature part becomes more pronounced. 

\begin{figure}
\centering
     \begin{subfigure}[]{0.49\textwidth}
         \centering
         \includegraphics[width=\textwidth]{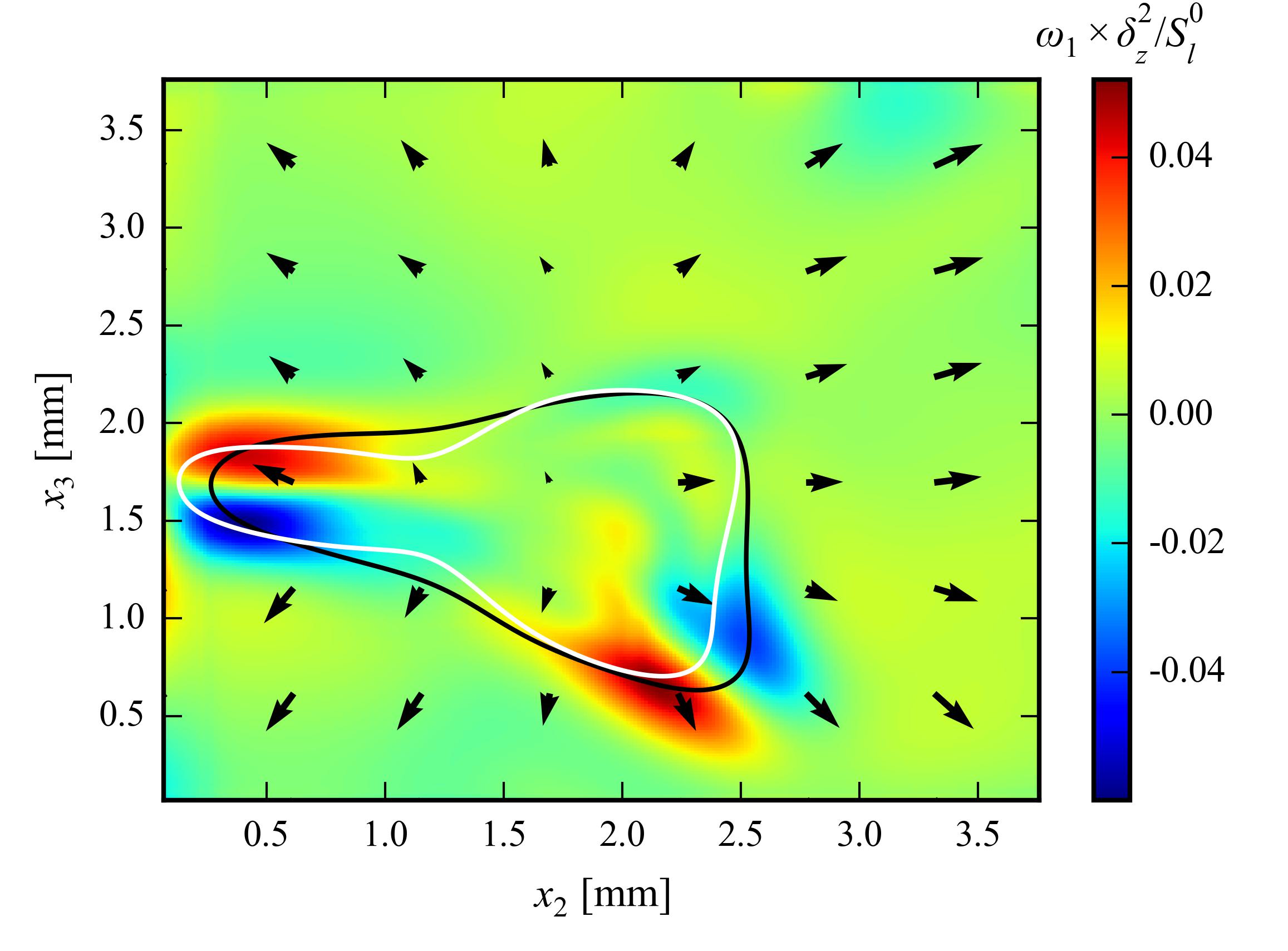}\vskip -8pt
         \caption{}
     \end{subfigure}
     \begin{subfigure}[]{0.49\textwidth}
         \centering
         \includegraphics[width=\textwidth]{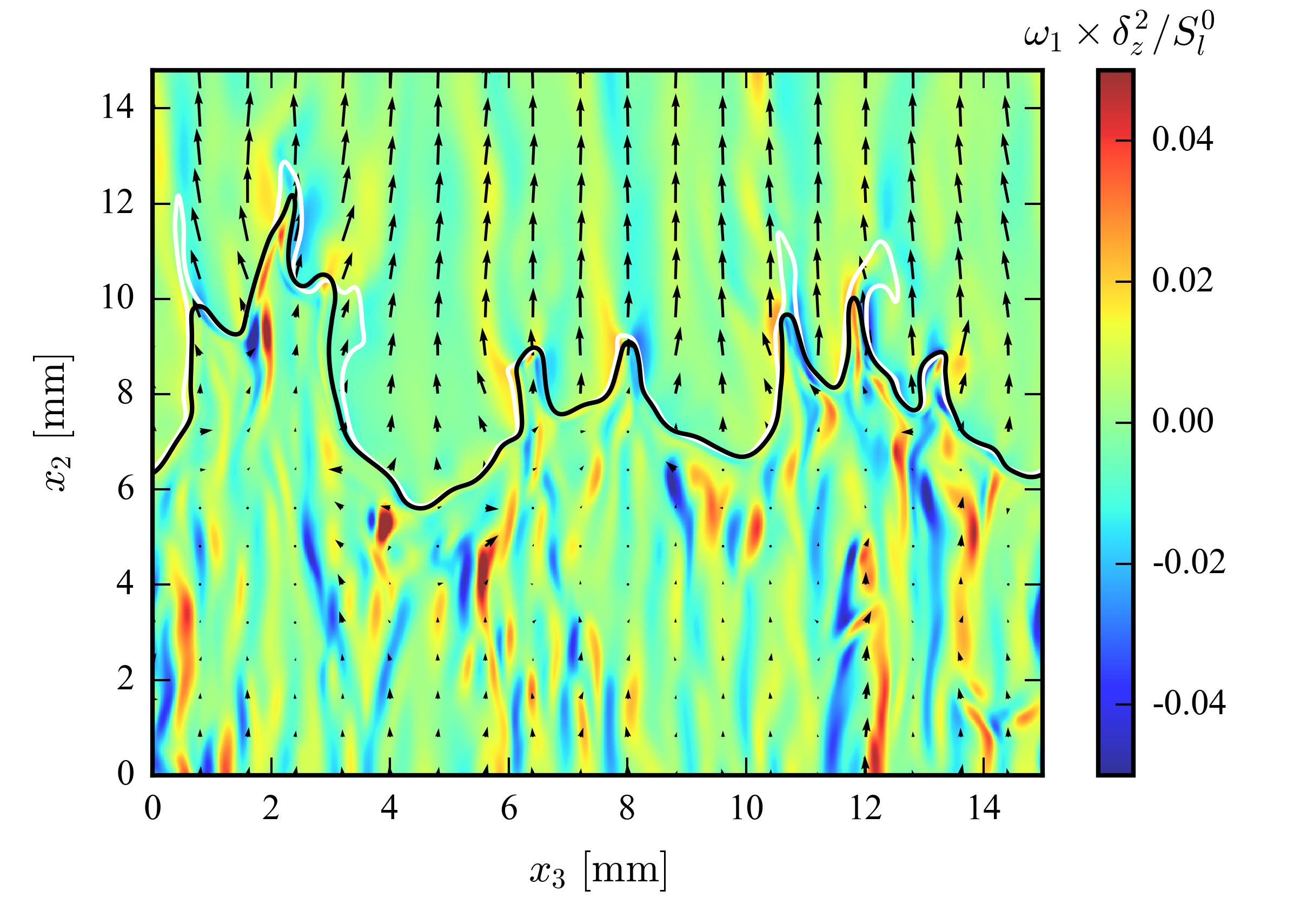}\vskip -8pt
         \caption{}
     \end{subfigure}
\caption{Contour plots of vorticity normal to the wall $\omega_1$ superimposed with the skin friction vector $\mathbf{F}$ (arrow size indicates its magnitude), where the black contour denote $c_{\mathrm{H}_2}=0.8$ contour line and the white contour represents the $c_{\mathrm{O}_2}=0.64$ contour line for different configurations: (a) wall normal flushing flame, (b) inclined sweeping flame.}
\label{fig:demoCurvOmega}
\end{figure}

The above joint PDF results can be well understood from flow visualization in Fig.~\ref{fig:demoCurvOmega}. The surface distribution of $\omega_1$ is superimposed with the skin friction vector, where the arrow size indicates its magnitude. The black and white lines denote the hydrogen contour line $c=0.8$ and oxygen contour line $c_{\mathrm{O}_2}=0.64$, respectively. For both cases, the oxygen contour lines exhibit larger curvature than the hydrogen contour lines. The same tendency also appears for the flame away from the wall~\citep{zhao2022near}. This property can be explained by the diffusion term of species $Y_i$, i.e., $\nabla\cdot(\rho D_i\nabla Y_i)$, which can be decomposed as
\begin{equation}
   \nabla\cdot(\rho D_i\nabla Y_i)=-\rho D_i \kappa|\nabla Y_i|+\boldsymbol{n}\cdot\nabla(\rho D_i\boldsymbol{n}\cdot\nabla Y_i).
   \label{diff}
\end{equation}
In the flame zone, the mass consumption ratio between different species is overall stoichiometrically determined. Since the diffusion coefficient of $\mathrm{H}_2$ is larger than that of $\mathrm{O}_2$, then the $\mathrm{O}_2$ contour curvature needs to be larger to ensure sufficient diffusion from the curvature controlled part $\rho D_i \kappa|\nabla Y_i|$ to balance diffusion of $\mathrm{H}_2$ in premixed combustion. In the following, the progress variable $c$ by default is hydrogen defined. Interestingly, Fig.~\ref{fig:demoCurvOmega} shows the vortex pair structure, where a positive $\omega_1$ and a negative $\omega_1$ region are closely allocated. Moving from the weak vortex region into the strong vortex region, either positive or negative, the contour line curvature starts to increase, till the curvature reaches maximum (magnitude) somewhere between a vortex pair. In summary, such a mechanism, consistent with the statistical joint PDFs in Fig.~\ref{fig:CurOmegaH2} and~\ref{fig:CurOmegaO2}, is a natural outcome of the flame and velocity interaction on the wall. 

It is generally believed that in turbulence the scalar gradient tends to align with the most compressive direction in space~\citep{ashurst1987alignment}. With reference to the boundary free flame property, effects of the flame normal strain rate $S_n=n_in_j\mathbf{S}_{ij}$ and flame tangential strain rate $S_t=(\delta_{ij}-n_in_j)\mathbf{S}_{ij}$ ($\delta_{ij}$ is the Kronecker delta) are also discussed.

\begin{figure}
\centering
     \begin{subfigure}[]{0.49\textwidth}
         \centering
         \includegraphics[width=\textwidth]{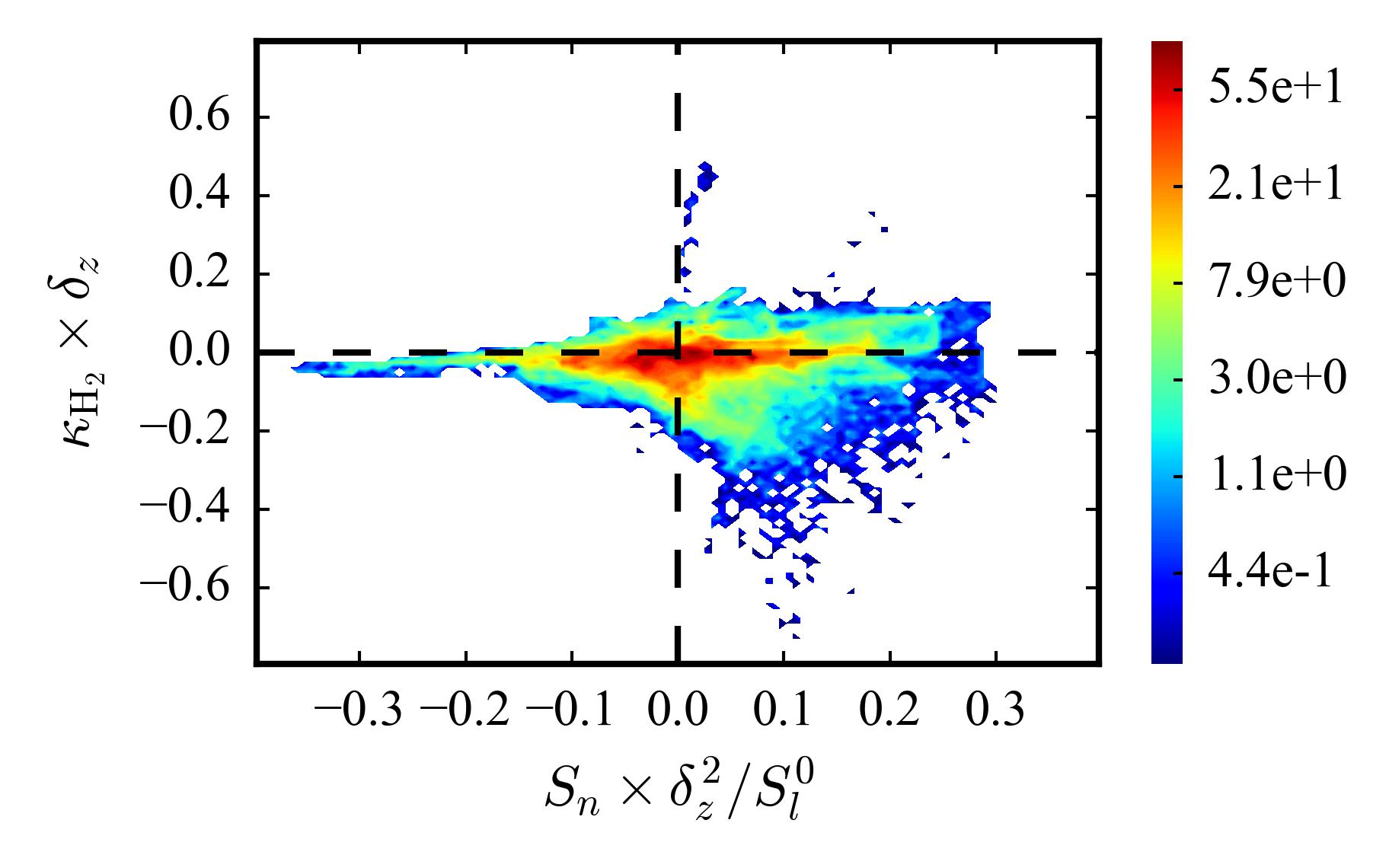}\vskip -8pt
         \caption{}
     \end{subfigure}
     \begin{subfigure}[]{0.49\textwidth}
         \centering
         \includegraphics[width=\textwidth]{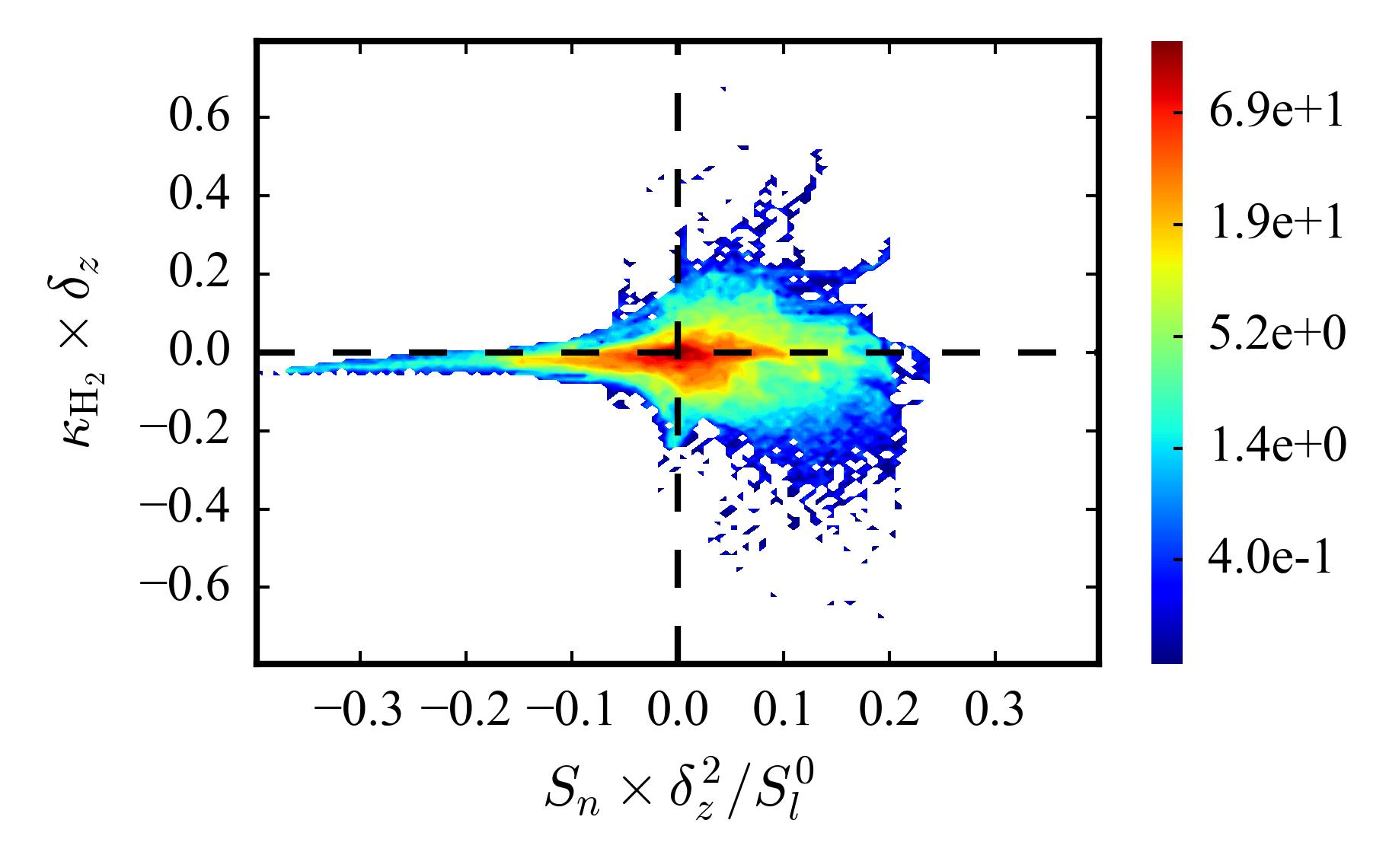}\vskip -8pt
         \caption{}
     \end{subfigure}
     \begin{subfigure}[]{0.49\textwidth}
         \centering
         \includegraphics[width=\textwidth]{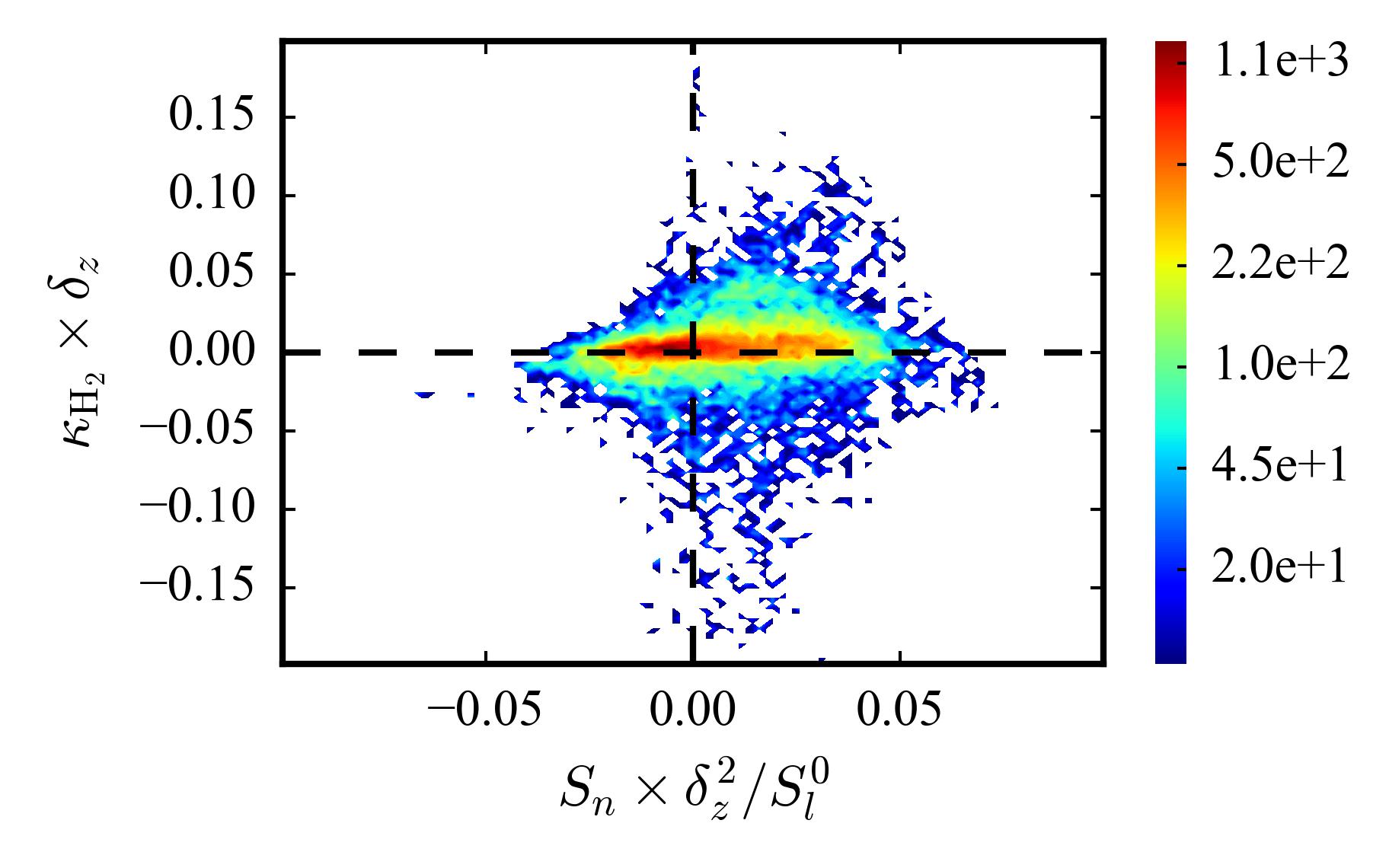}\vskip -8pt
         \caption{}
     \end{subfigure}
     \begin{subfigure}[]{0.49\textwidth}
         \centering
         \includegraphics[width=\textwidth]{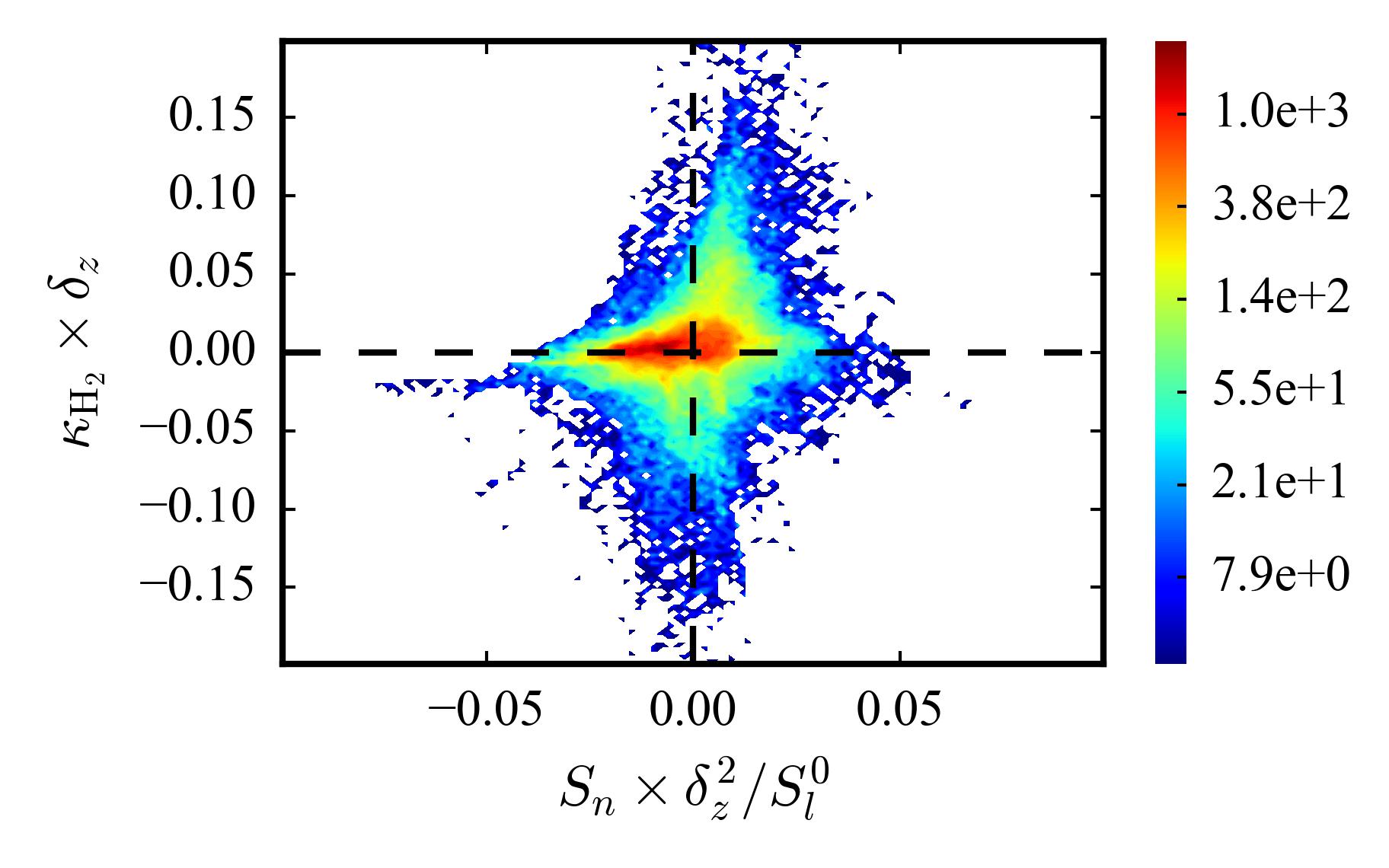}\vskip -8pt
         \caption{}
     \end{subfigure}
\caption{Joint PDFs between curvature $\kappa_{\mathrm{H}_2}$ and flame normal strain rate $S_n$ for normal flushing flame: (a) $0.65< c\leq0.8$; (b) $0.8 <c\leq0.95$; and for inclined sweeping flame: (c) $0.65 <c\leq0.8$; (d) $0.8< c\leq0.95$.}
\label{fig:CurSnH2}
\end{figure}

\begin{figure}
\centering
     \begin{subfigure}[]{0.49\textwidth}
         \centering
         \includegraphics[width=\textwidth]{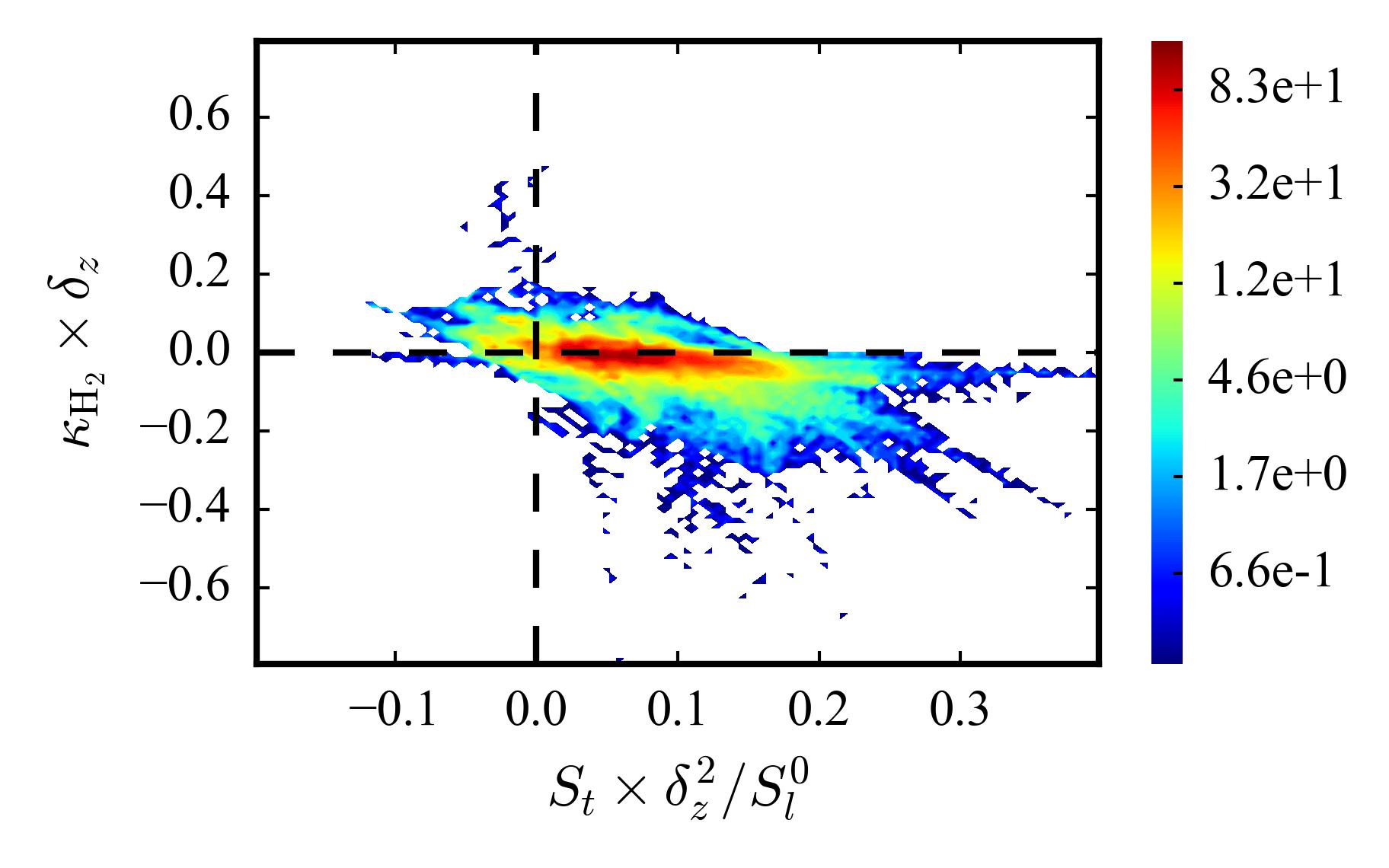}\vskip -8pt
         \caption{}
     \end{subfigure}
     \begin{subfigure}[]{0.49\textwidth}
         \centering
         \includegraphics[width=\textwidth]{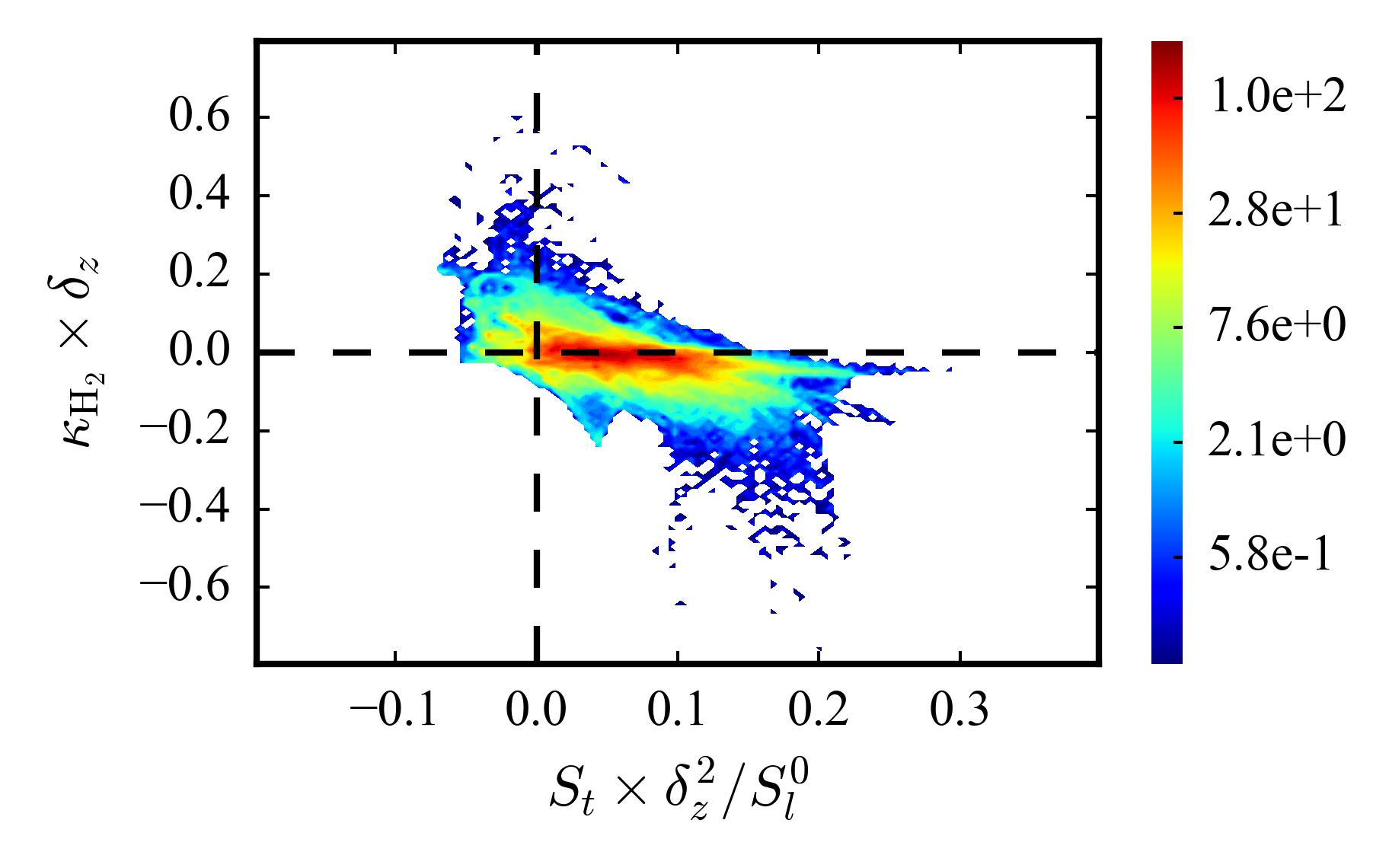}\vskip -8pt
         \caption{}
     \end{subfigure}
     \begin{subfigure}[]{0.49\textwidth}
         \centering
         \includegraphics[width=\textwidth]{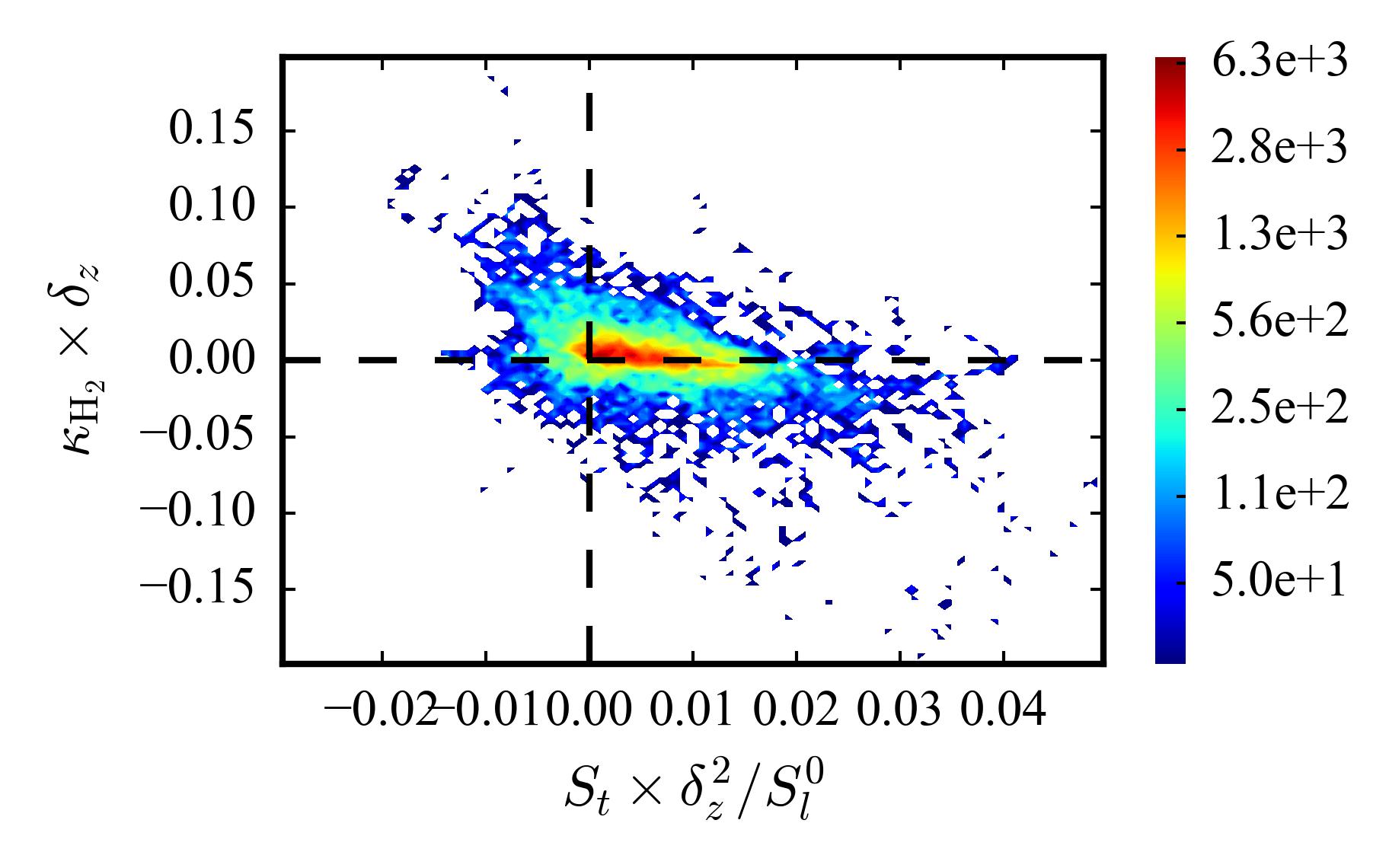}\vskip -8pt
         \caption{}
     \end{subfigure}
     \begin{subfigure}[]{0.49\textwidth}
         \centering
         \includegraphics[width=\textwidth]{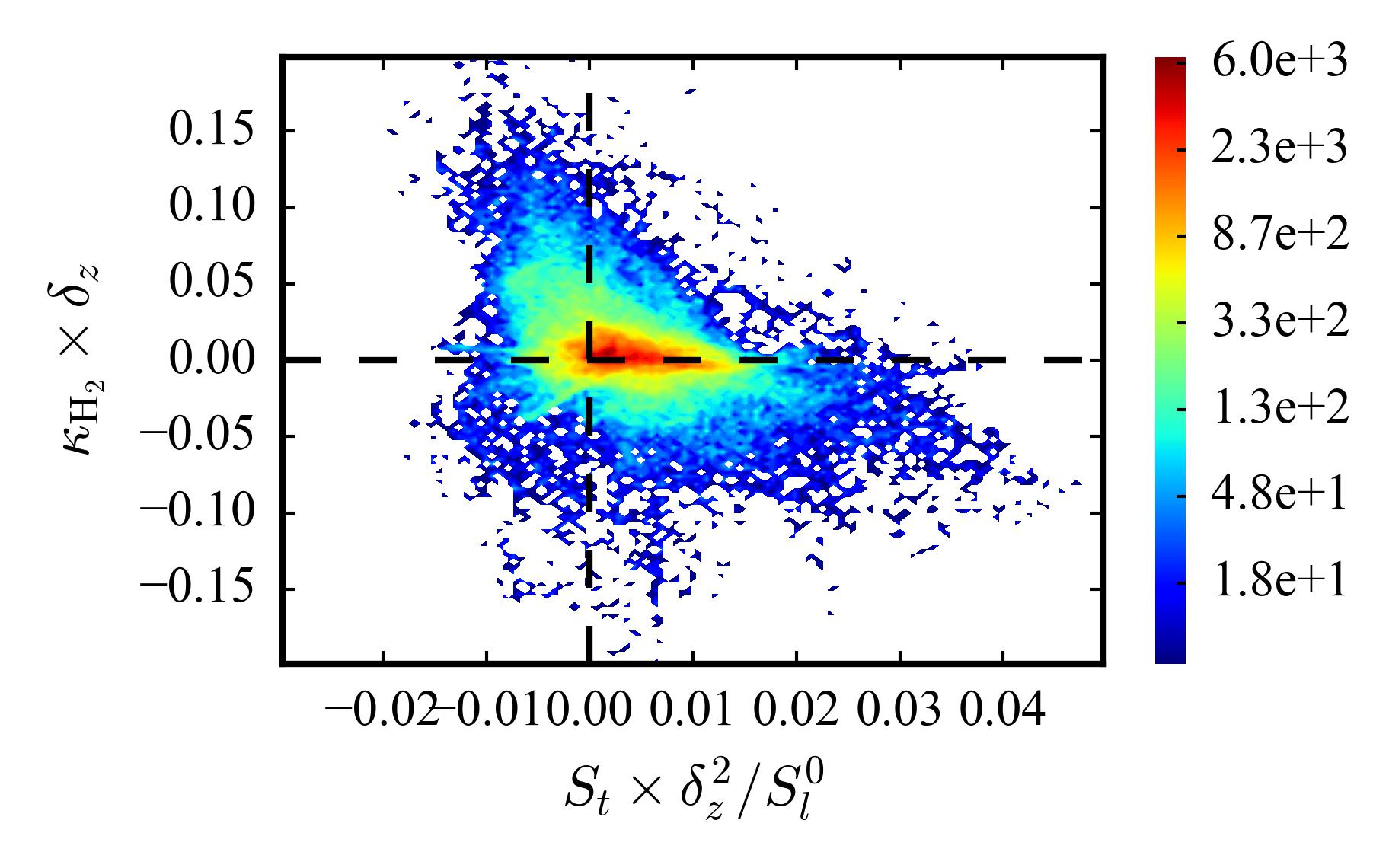}\vskip -8pt
         \caption{}
     \end{subfigure}
\caption{Joint PDFs between curvature $\kappa_{\mathrm{H}_2}$ and flame tangential strain rate $S_t$ for the normal flushing flame conditional on: (a) $0.65< c\leq0.8$; (b) $0.8 <c\leq0.95$, and inclined sweeping flame conditional on: (c) $0.65 <c\leq0.8$; (d) $0.8< c\leq0.95$.}
\label{fig:CurStH2}
\end{figure}

From the normal strain rate results in Fig.~\ref{fig:CurSnH2}, it can be seen that because of thermal expansion, the positive $S_n$ is predominant, especially in Fig.~\ref{fig:CurSnH2} (a), (b), and (c). Meanwhile, it is worth noting that on the wall boundary with constant temperature, such expansion effect is much weakened, compared with the boundary free results. For samples with large curvature magnitude, the normal strain rate $S_n$ is close to zero. Consistently when $S_n$ is strongly negative, the flame front is likely flat with almost zero curvature. Dependence of the flame curvature on the flame tangential strain rate is shown in Fig.~\ref{fig:CurStH2}. Overall, there exists a weakly negative correlation, i.e., the flame curvature is smaller if the flame is more stretched, while the curvature tends to be strongly positive under the action of compressive tangential strain rate. 

\begin{figure}
\centering
     \begin{subfigure}[]{0.49\textwidth}
         \centering
         \includegraphics[width=\textwidth]{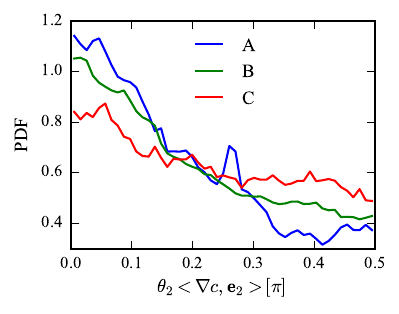}\vskip -8pt
         \caption{}
     \end{subfigure}
     \begin{subfigure}[]{0.49\textwidth}
         \centering
         \includegraphics[width=\textwidth]{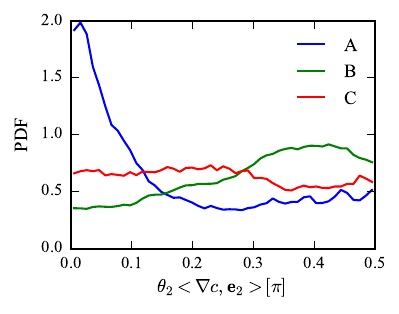}\vskip -8pt
         \caption{}
     \end{subfigure}
\caption{PDFs of the acute angle $\theta_2$ between the progress variable gradient $\nabla c$ and the eigenvector $\mathbf{e}_2$ conditioned in different regions for: (a) wall normal flushing flame; (b) inclined sweeping flame.}
\label{fig:eigenAngle}
\end{figure}

For more statistical details, the wall surface is partitioned into three regions ($A$, $B$, and $C$) based on the progress variable gradient magnitude, with the non-reactive region (if the wall heat flux $Q_w<10^{-5}$) excluded. Specifically, region $A$, $B$, and $C$ correspond to high gradient (top $10\%$ samples), mediate gradient ($80\%$ samples in the middle), and low gradient region (bottom $10\%$ samples), respectively. $\theta_2$ is defined as the acute angle between the progress variable gradient $\nabla c$ and the most compressive eigenvector $\mathbf{e}_2$ of the tensor $\{\mathbf{S}_{ij}\}$. By definition, $\theta_2=0$ means $\nabla c$ aligns perfectly with the most compressive strain direction, while $\theta_2=\pi/2$ means a perfect alignment between $\nabla c$ and the most extensive strain direction. As shown in Fig.~\ref{fig:eigenAngle},  for both cases there is a strong alignment between $\nabla c$ and $\mathbf{e}_2$ in region A. Moving away from the flame zone with decreasing progress variable gradient magnitude, such alignment weakens.

\subsection{Species alignment relation}
Wall interference leads to the near-wall flame structure largely different from that of the boundary free flame. In the present analysis, we focus on the alignment relation among different species isosurfaces. If the flame is infinitely thin as a limiting case, the normal directions of all these isosurfaces are the same as the flame normal, which, however, can not be expected for the near-wall flame. We tentatively introduce a newly defined alignment index as 
\begin{equation}
    G_{\mathrm{align}} = \frac{\sum_{P(i,j)} |\nabla Y_i\cdot \nabla Y_j|}{\sum_{P(i,j)} |\nabla Y_i| |\nabla Y_j|}.
\end{equation}
Here $P(i,j)$ represents the pairing of any two species $i$ and $j$, whose contribution to the alignment among all the species is weighted by the gradient magnitude $\nabla Y_i$ and $\nabla Y_j$. Such a weighting factor is effective to exclude the noisy influence from minor species. Clearly, if all species (anti-)align perfectly, $G_{\mathrm{align}} = 1$, while for purely random structure, analytically $G_{\mathrm{align}}$ is $0.5$.

\begin{figure}
     \begin{subfigure}[]{0.49\textwidth}
        \centering
    \includegraphics[width=\textwidth]{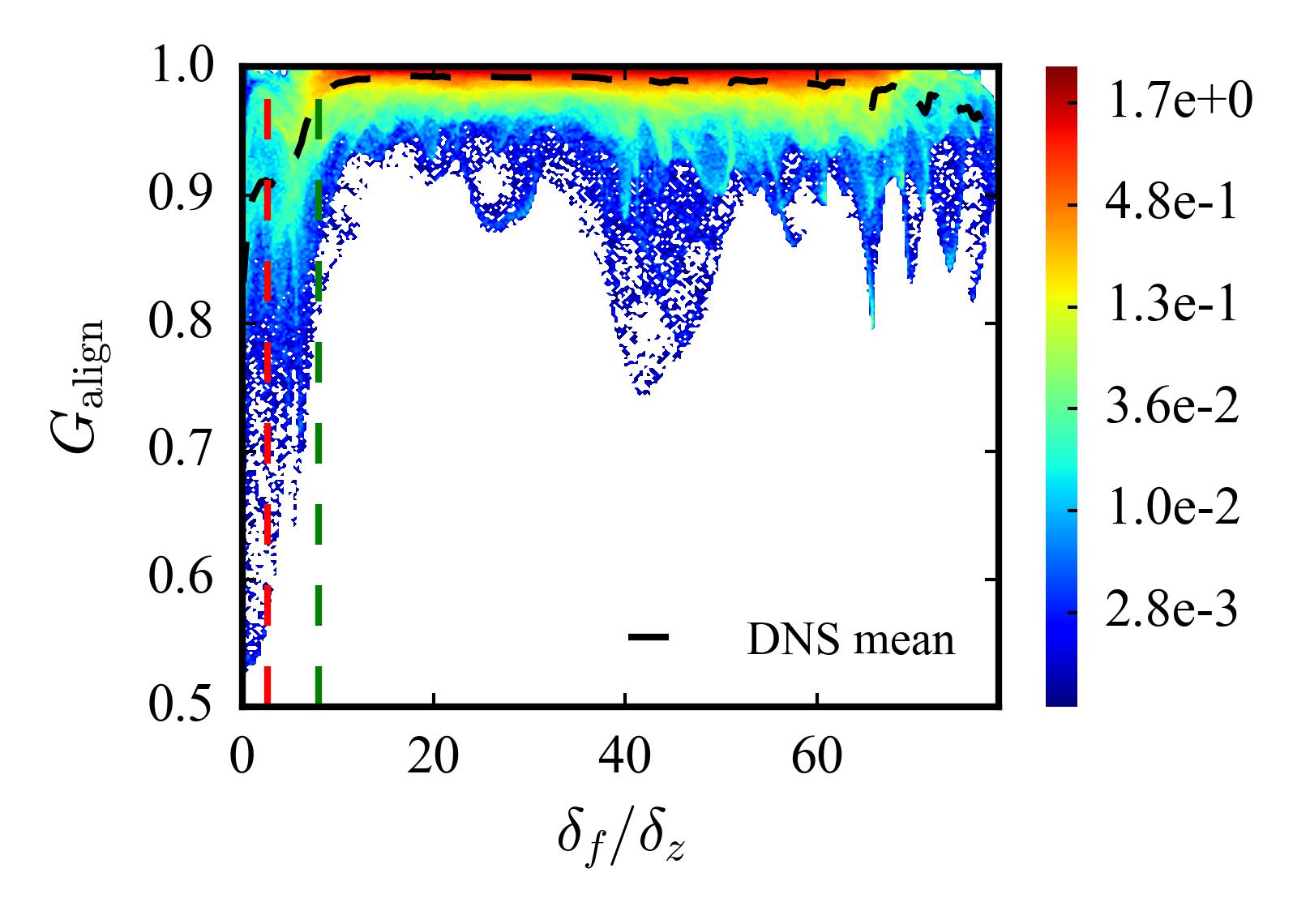}\vskip -4pt
         \caption{}
     \end{subfigure}
     \begin{subfigure}[]{0.49\textwidth}
        \centering
        \includegraphics[width=\textwidth]{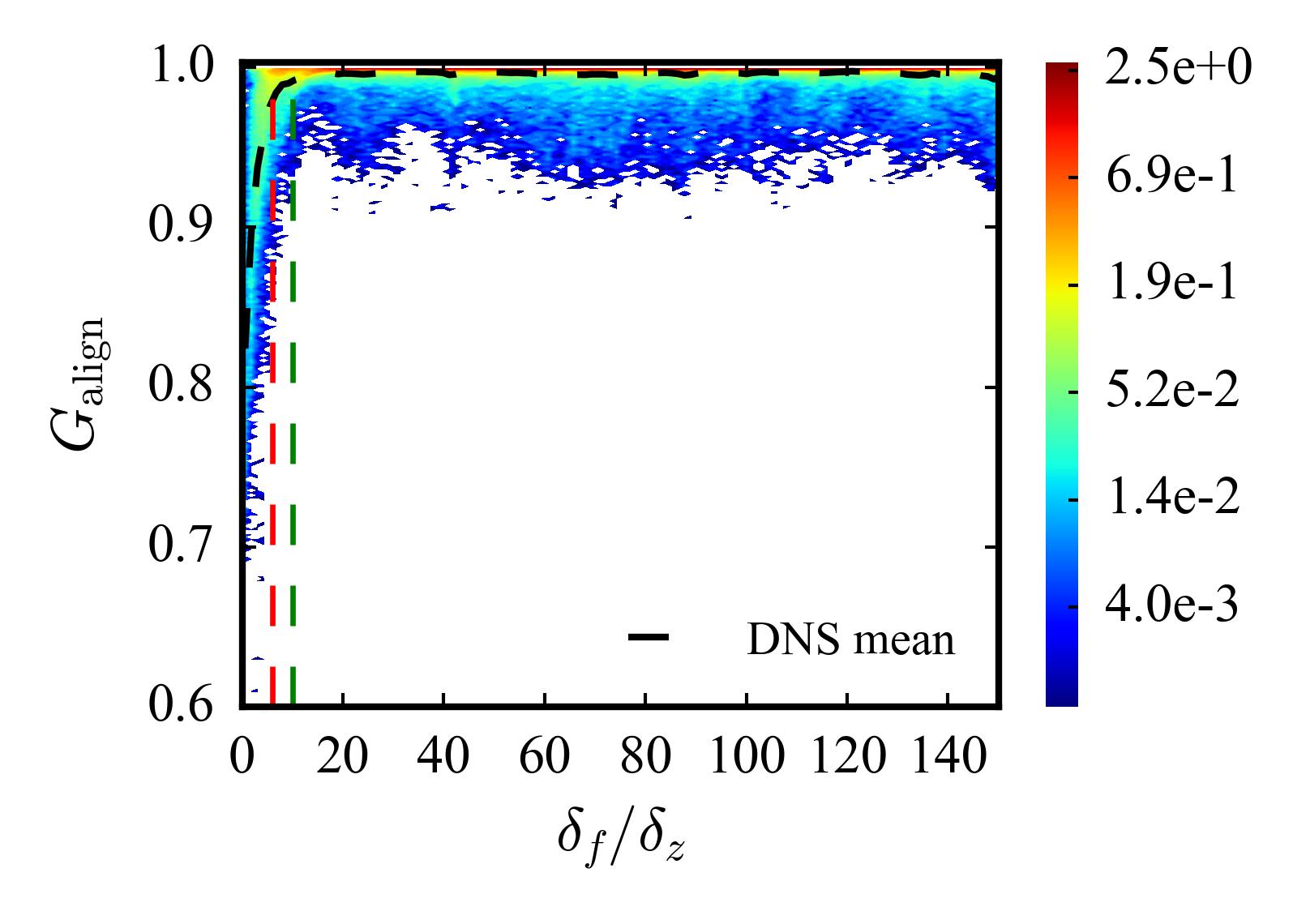}\vskip -4pt
         \caption{}
     \end{subfigure}
    \caption{Joint PDFs between the flame-wall distance and flame alignment index $G_{\mathrm{align}}$, for: (a) wall normal flushing flame; (b) inclined sweeping flame. The vertical red and green dashed lines mark the quenching zone and the influenced zone, respectively.}
    \label{fig:Galign08}
\end{figure}

Fig.~\ref{fig:Galign08} shows for both configurations the joint PDF between $G_{\mathrm{align}}$ and the flame-wall distance $\delta_f$ in different zones. For the wall normal flushing flame, the quenching zone, influence zone, and free flame zone are quantified as $\delta_f/\delta_z < 2.66$, $2.66
\leq\delta_f/\delta_z\leq 8$ and $\delta_f/\delta_z > 8$, respectively~\citep{2018Analysis}. For the inclined sweeping flame setup with a much lower wall normal strain rate, the quenching zone and influence zone are differently quantified as $\delta_f/\delta_z < 6.0$~\citep{bellenoue2003direct,jainski2017sidewall,kosaka2018wall} and $6\leq\delta_f/\delta_z\leq 10$, respectively. It is seen that consistently the mean alignment index approaches $1.0$ in the free flame zone. Moving towards the wall to the influence zone (green dashed line) and quenching zone (red dashed line), the flame becomes thicker with more pronounced differential diffusion. Consequently, $G_{\mathrm{align}}$ continues to decrease. 

\begin{figure}
     \begin{subfigure}[]{0.49\textwidth}
        \centering
        \includegraphics[width=\textwidth]{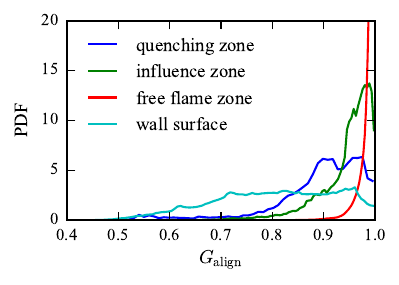}\vskip -4pt
         \caption{}
     \end{subfigure}
     \begin{subfigure}[]{0.49\textwidth}
        \centering
        \includegraphics[width=\textwidth]{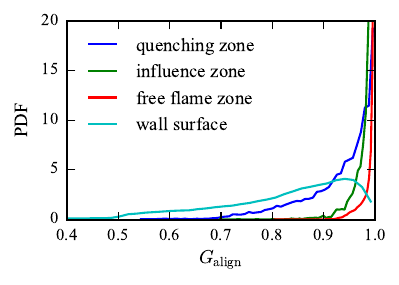}\vskip -4pt
         \caption{}
     \end{subfigure}
     \caption{PDFs of the flame alignment index $G_{\mathrm{align}}$ in different zones, including the wall surface results, for: (a) wall normal flushing flame, (b) inclined sweeping flame.}
    \label{fig:GalignAll}
\end{figure}

In more detail, PDFs of $G_{\mathrm{align}}$ are further examined. Figure~\ref{fig:GalignAll} shows that as moving towards the wall, the PDF is more distributed with a decreasing expectation, which implies that the flame structure becomes more chaotic if the wall influence intensifies. Especially on the wall boundary, where species convection and reactions become negligible, the dominant differential diffusion leads to stronger species misalignment. The small difference between Fig.~\ref{fig:GalignAll} (a) for the wall normal flushing flame and Fig.~\ref{fig:GalignAll} (b) for the inclined sweeping flame may arise from the different species transport models, i.e., mixture-averaged transport model in Fig.~\ref{fig:GalignAll} (a) and constant Lewis number assumption in Fig.~\ref{fig:GalignAll} (b). Because of such imperfect species alignment relation, the movement of different $c$-isosurfaces near the wall is inconsistent, which then generates complex thickened flames. In such a sense, it is not suitable to use the conventional flame speed (e.g., displacement speed or the diffusion speed) for a certain isosurface to describe the thickened flame movement. Instead, the so-called flame zone speed $S_{\mathrm{zone}}$ is appropriately introduced~\citep{2018Analysis} as
\begin{equation}
S_{\mathrm{zone}} = \frac{\int S_d {\rho}|{\nabla} c| \mathrm{d} \boldsymbol{n}}{{\rho}_u \int\left|{\nabla} c\right| \mathrm{d} \boldsymbol{n}}.
\end{equation}
Here $S_{\mathrm{zone}}$ represents the ${\nabla} c$ weighted displacement speed $S_d$ along the flame normal across the entire flame zone.

\subsection{Near-wall heat transfer}
\begin{figure}
     \begin{subfigure}[]{0.49\textwidth}
        \centering
        \includegraphics[width=\textwidth]{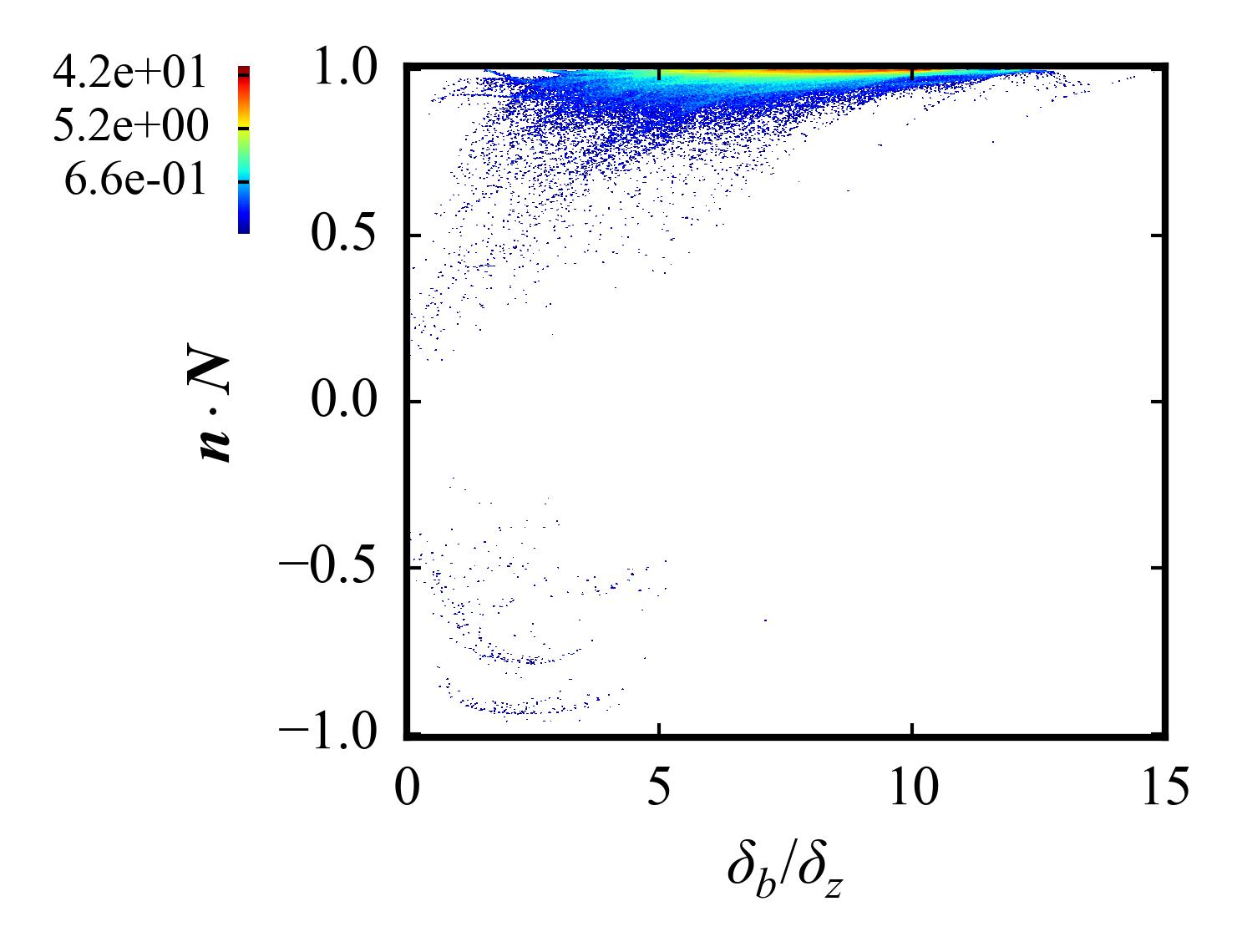}\vskip -4pt
         \caption{}
     \end{subfigure}
     \begin{subfigure}[]{0.49\textwidth}
        \centering
        \includegraphics[width=\textwidth]{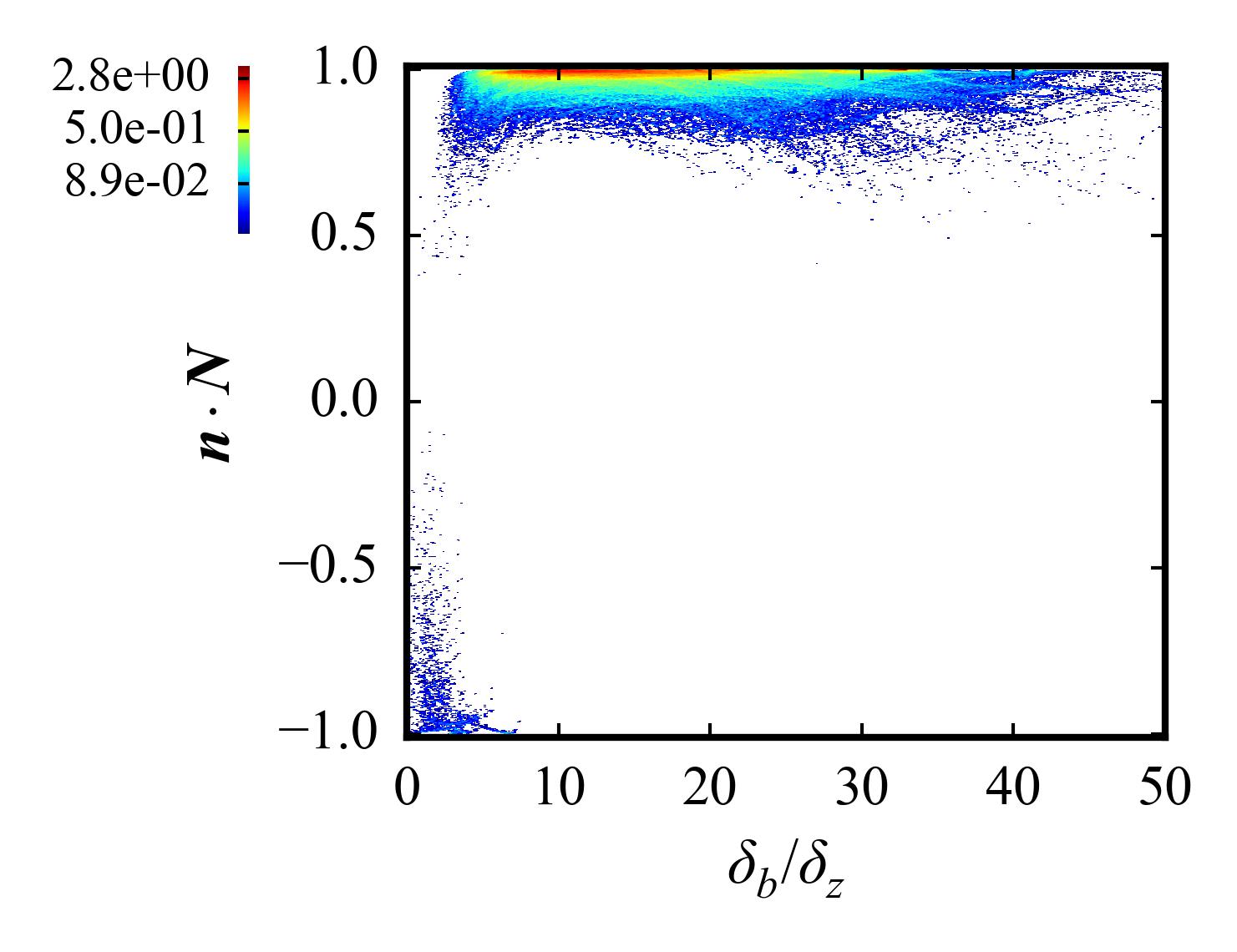}\vskip -4pt
         \caption{}
     \end{subfigure}
    \caption{Joint PDFs between $\textbf{\textit{n}}\cdot\textbf{\textit{N}}$ and flame burnt boundary to wall distance $\delta_b/ \delta_z$, for both configurations: (a) wall normal flushing flame, (b) inclined sweeping flame.}
    \label{fig:cos}
\end{figure}

According to the relative orientation between the flame normal $\textbf{\textit{n}}$ and the wall normal $\textbf{\textit{N}}$ (pointing toward the fluid), the near-wall flame can be classified~\citep{2018Analysis} as head-on, if $\textbf{\textit{n}}\cdot\textbf{\textit{N}}> 0$ or entrained if $\textbf{\textit{n}}\cdot\textbf{\textit{N}}< 0$. Fig.~\ref{fig:cos} shows that the dominant part of both flow setups is head-on with $\textbf{\textit{n}}\cdot\textbf{\textit{N}}$ close to 1. Even for the inclined sweeping flame in stronger turbulence, the entrained part is still negligibly small. Therefore, in the following we focus on the head-on flame analysis.

\begin{figure}
     \begin{subfigure}[]{0.49\textwidth}
        \centering
        \includegraphics[width=0.99\textwidth]{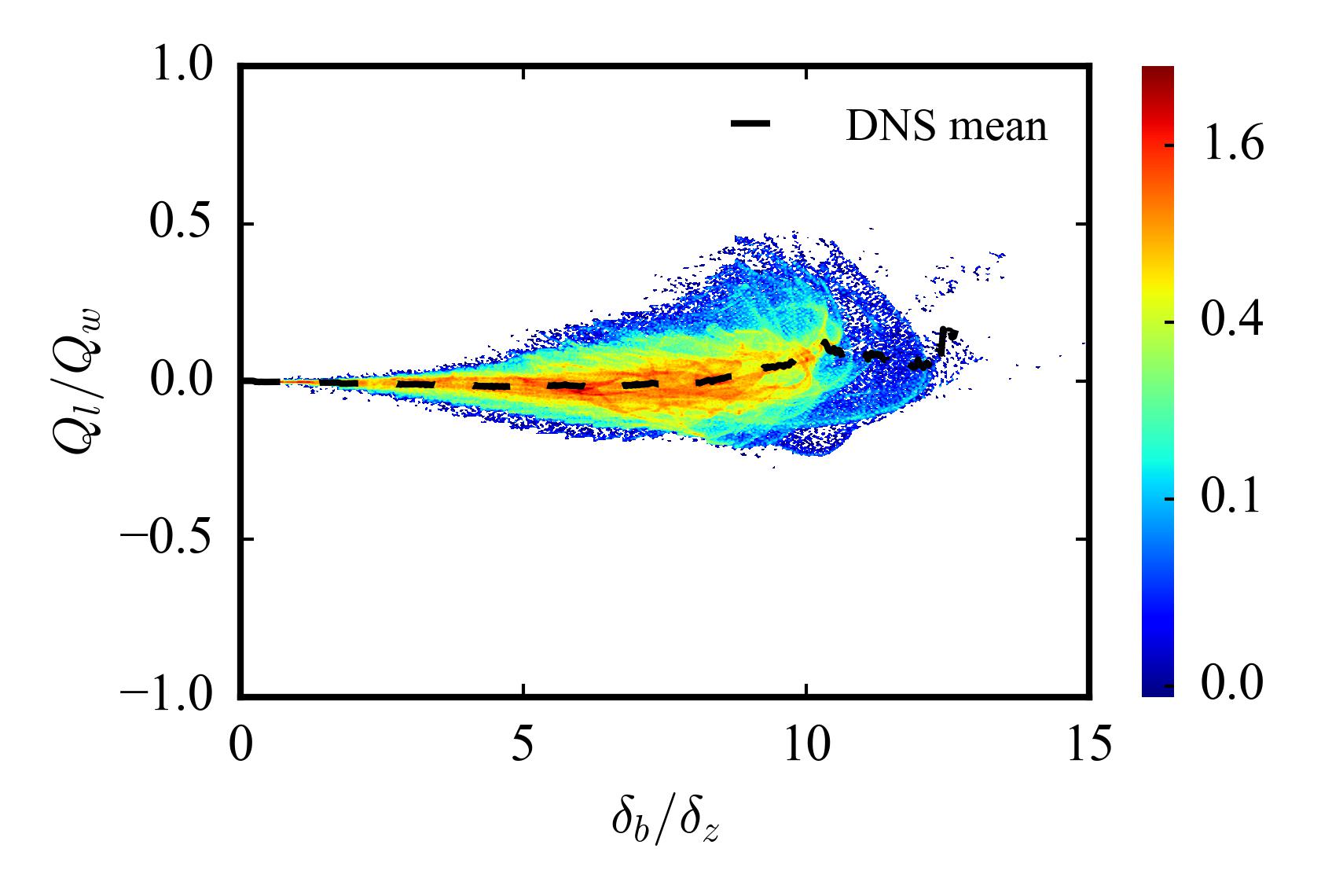}\vskip -4pt
         \caption{}
     \end{subfigure}
     \begin{subfigure}[]{0.49\textwidth}
        \centering
        \includegraphics[width=\textwidth]{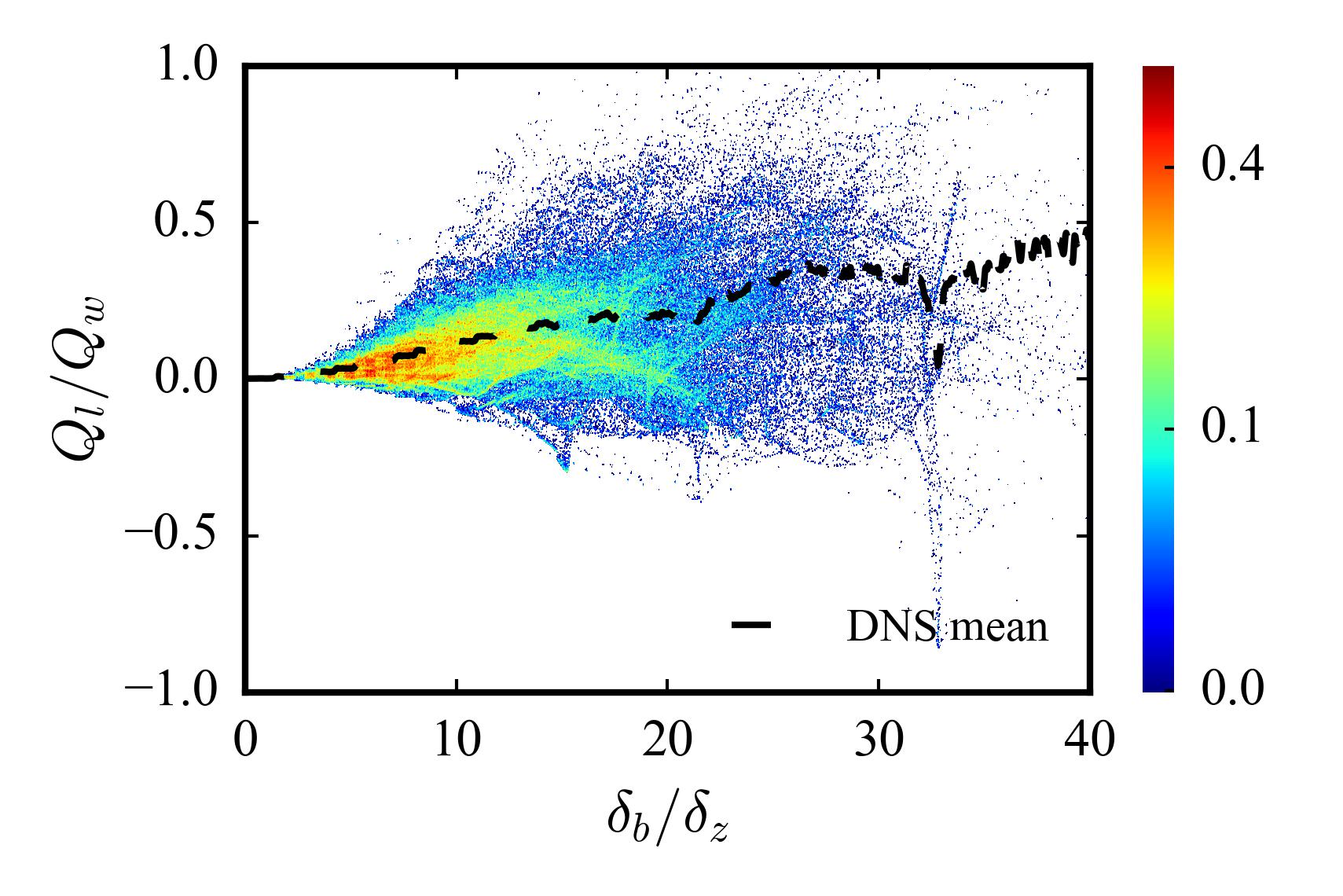}\vskip -4pt
         \caption{}
     \end{subfigure}
    \caption{Relative importance of the lateral heat transfer $Q_l$ to $Q_w$ with respect to $\delta_b/ \delta_z$ for: (a) wall normal flushing flame; (b) inclined sweeping flame.}
    \label{fig:ql}
\end{figure}

Since statistically the flame normal aligns with the wall normal, it is expected that thermal energy transfer along the wall normal, including the diffusion term $q_{n,\mathrm{diff}} = \frac{1}{C_p}\frac{1}{RePr} \frac{\partial \left( \lambda \frac{\partial T}  {\partial x_1} \right) } {\partial x_1}$ and convection term $q_{n,\mathrm{conv}} = \rho u_1 \frac{\partial T}{\partial x_1}$, dominates over the wall parallel part $q_l$
\begin{equation}
q_l = \frac{1}{C_p}\frac{1}{RePr} \left[\frac{\partial \left( \lambda \frac{\partial T}  {\partial x_2} \right) } {\partial x_2} + \frac{\partial \left( \lambda \frac{\partial T}  {\partial x_3} \right) } {\partial x_3} \right]- \rho u_2 \frac{\partial T}{\partial x_2} - \rho u_3 \frac{\partial T}{\partial x_3}.
\label{Eq:ql}
\end{equation}
To quantify the influence of $q_l$ on each flame element, we consider within the burnt region the integrals 
$\int_{-\delta_b}^{0} q_{n,\mathrm{diff}} \mathrm{d}x_1$, $\int_{-\delta_b}^{0} q_{n,\mathrm{conv}} \mathrm{d}x_1$ and $\int_{-\delta_b}^{0} q_l \mathrm{d}x_1$. Since $\int_{-\delta_b}^{0} q_{n,\mathrm{diff}} \mathrm{d}x_1\sim\frac{Q_w}{Cp}$, it is meaningful to check the ratio between $Q_l = \int_{-\delta_b}^{0} C_p q_l \mathrm{d}x_1$ and $Q_w$. As presented in Fig.~\ref{fig:ql} (a) and (b), because of the isothermal wall conditions, $Q_l/Q_w$ vanishes as $\delta_b$ approaches zero. Away from the wall, although the magnitude slightly increases, on average $Q_l/Q_w$ remains negligibly small. For the inclined sweeping flame results in Fig.~\ref{fig:ql} (b), because at the large scale there exists a flame front inclination angle, $Q_l/Q_w$ increases with $\delta_b$, but overall is still small.

\section{Wall heat flux modeling and validation}
\label{S2}
In view of the practical importance of the wall heat flux quantity, a predictive model is proposed based on the obtained fundamental properties.

\subsection{Model construction}
\begin{figure}
    \centering
    \includegraphics[width=0.7\textwidth]{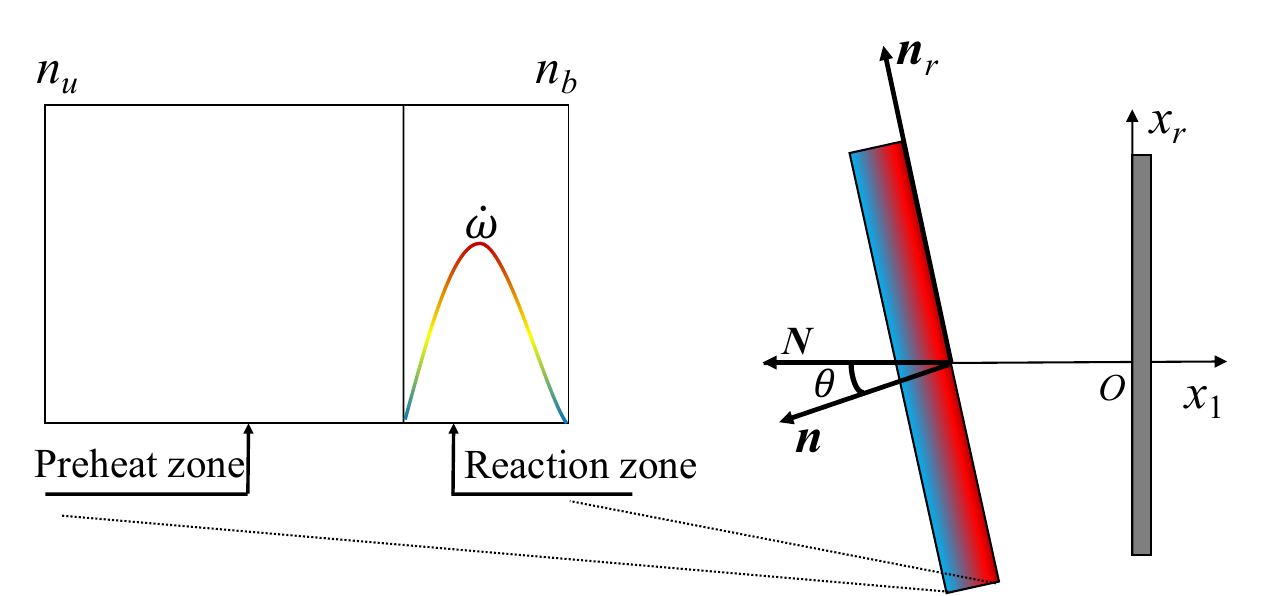}\label{fig:headon}
    \caption{Schematic of simplified FWI scenarios for the head-on flame case.}
    \label{fig:schematic}
\end{figure}

In the vicinity of the wall boundary at $x_1=0$, a local piece of the flame front is shown in Fig.~\ref{fig:schematic}. The flame segment consists of a preheat zone and a reaction zone ranging from $n_{u}$ to $n_{b}$, inclined with an angle $\theta$ relative to the wall.

In the following, the model construction is introduced concerning the temperature field and the velocity field, respectively.

\subsubsection{Temperature field}
The simplified temperature and progress variable equations in non-dimensional form are listed as follows:
\begin{equation}
    {\rho}\frac{\partial {T}}{\partial {t}} + {\rho} {\textbf{\textit{u}}} \cdot {\nabla} {T} = \frac{1}{C}_p\frac{1}{RePr} {\nabla} \cdot \left({\lambda} {\nabla}{T}\right) + \frac{{\Dot{\omega}}_T}{{C}_p},
\label{T}
\end{equation}
\begin{equation}
    {\rho}\frac{\partial c}{\partial {t}} + {\rho} {\textbf{\textit{u}}} \cdot {\nabla} c = \frac{1}{ReSc} {\nabla} \cdot \left({\rho}{D} {\nabla}c \right) + {\Dot{\omega}}_c.
\label{C}
\end{equation}

Integrating Eq.~\eqref{T} and Eq.~\eqref{C} from the burnt side $n_{b}$ to the unburnt side $n_{u}$ along the local flame normal direction $\textbf{\textit{n}}$, we obtain
\begin{equation}
    \int_{n_{b}}^{n_{u}} {\rho}\frac{\partial  {T}}{\partial  {t}} \, \mathrm{d} n + \dot{m}_T \int_{n_{b}}^{n_{u}} {\nabla} {T} \cdot \mathrm{d}\textbf{\textit{n}} = \int_{n_{b}}^{n_{u}} \frac{1}{{C}_p}\frac{1}{RePr} {\nabla} \cdot \left({\lambda} {\nabla}{T}\right) \mathrm{d} n + \int_{n_{b}}^{n_{u}} {\frac{\Dot{\omega}_T}{{C}_p} } \, \mathrm{d} n,
\label{intT}
\end{equation}

\begin{equation}
    \int_{n_{b}}^{n_{u}} {\rho}\frac{\partial  {c}}{\partial  {t}} \, \mathrm{d} n + \dot{m}_c \int_{n_{b}}^{n_{u}} {\nabla} {c} \cdot \mathrm{d}\textbf{\textit{n}} = \int_{n_{b}}^{n_{u}} \frac{1}{ReSc} {\nabla} \cdot \left({\rho}{D} {\nabla}c \right)  \mathrm{d} n + \int_{n_{b}}^{n_{u}} {\Dot{\omega}_c} \, \mathrm{d} n.
\label{intC}
\end{equation}
Here a temperature based mean mass flux across the flame zone $\dot{m}_T \equiv \int_{n_{b}}^{n_{u}} \rho \textbf{\textit{u}}\cdot \nabla T \, \mathrm{d} n / {\int_{n_{b}}^{n_{u}} \nabla T \cdot \mathrm{d}\textbf{\textit{n}}}$, and in parallel a progress variable based mean mass flux $\dot{m}_c \equiv \int_{n_{b}}^{n_{u}} \rho \textbf{\textit{u}}\cdot \nabla c \, \mathrm{d} n / {\int_{n_{b}}^{n_{u}} \nabla c \cdot \mathrm{d}\textbf{\textit{n}}}$. In model implementation, $\dot{m}_T$ and $\dot{m}_c$ are in-situ input parameters. Typically $\dot{m}_T=\dot{m}_c$, for instance, a freely propagating thin flame. However, for the near-wall thickened flame with complex structures, these two mass fluxes are not necessarily identical.

In the unburnt upstream region, the temperature and the progress variable remain constant as the fresh incoming condition, for instance, the temperature at $n_{u}$, i.e., $T_u = 0$ and the progress variable at $n_{u}$, i.e., $c_u = 0$. In the region between the wall and the burnt side of the flame, the gradient of the species mass fraction is negligible since the weak reactions in the region and the zero-gradient condition on the wall. Therefore, Eq.~\eqref{intT} and Eq.~\eqref{intC} can be written as
\begin{equation}
    \int_{n_{b}}^{n_{u}} {\rho}\frac{\partial  {T}}{\partial  {t}} \, \mathrm{d} n - \dot{m}_T T_{b} = -\frac{1}{\overline{C}_p}\frac{1}{RePr} \left.  \left( \lambda \frac{\partial T}{\partial n} \right) \right |_{n_{b}} + \int_{n_{b}}^{n_{u}} {\frac{\Dot{\omega}_T}{{C}_p}} \, \mathrm{d} n,
\label{intTBC}
\end{equation}
\begin{equation}
    \int_{n_{b}}^{n_{u}} {\rho}\frac{\partial  {c}}{\partial  {t}} \, \mathrm{d} n - \dot{m}_c c_{b} = \int_{n_{b}}^{n_{u}} {\Dot{\omega}_c} \, \mathrm{d} n,
\label{intCBC}
\end{equation}
where $T_b$ is the temperature at $n_{b}$ and $c_b$ is the progress variable at $n_{b}$; $\overline{C}_p$ represents a mean $C_p$ in the flame zone. Combining Eq.~\eqref{intTBC} and Eq.~\eqref{intCBC} to eliminate the reaction source, one obtains
\begin{equation}
    \int_{n_{b}}^{n_{u}}  ({\rho}\frac{\partial  {T}}{\partial  {t}} -  {\rho}\frac{\partial  {c}}{\partial  {t}}) \mathrm{d} n - \dot{m}_T T_{b} = -\frac{1}{\overline{C}_p}\frac{1}{RePr} \left.  \left( \lambda \frac{\partial T}{\partial n} \right) \right |_{n_{b}} - \dot{m}_c c_{b}.
\end{equation}
The temperature gradient typically aligns with the flame normal, i.e., $\cfrac{\partial T}{\partial n} = -\cfrac{\partial T}{\partial \textit{x}_1} \cfrac{1}{\cos{\theta}}$. Thus, the above equation for a local flame element becomes 
\begin{equation}
     \dot{m}_T T_{b}-\int_{n_{b}}^{n_{u}}  ({\rho}\frac{\partial  {T}}{\partial  {t}} -  {\rho}\frac{\partial  {c}}{\partial  {t}}) \mathrm{d} n = -\frac{1}{\overline{C}_p}\frac{1}{RePr} \left.  \left( \lambda \frac{\partial T}{\partial \textit{x}_1} \frac{1}{\cos{\theta}} \right) \right |_{x_{1,b}} + \dot{m}_c c_{b},
\label{headOnBc}
\end{equation}
where $x_{1,b}$ is the $x_1$ projection of $n_{b}$.

In the region between the wall and the burnt side of the flame, the absence of the chemical source is assumed due to its negligible contribution to the total chemical source. The temperature equation along the wall-normal direction $x_1$ is
\begin{equation}
    \rho\frac{\partial T}{\partial t} + \rho u_1 \frac{\partial T}{\partial x_1} = \frac{1}{C_p}\frac{1}{RePr}  \frac{\partial \left( \lambda \frac{\partial T}  {\partial x_1} \right) } {\partial x_1} + q_l, \quad \text{for } x_1 \geq x_{1,b},
\label{1dT}
\end{equation}
where $q_l$ is the wall parallel heat transfer term defined in Eq.~\eqref{Eq:ql}. Integrating the temperature equation from $x_{1,b}$ to the wall, we obtain

\begin{equation}
  \int_{x_{1,b}}^0 \rho\frac{\partial T}{\partial t} \, \mathrm{d} x_1 + \frac{1}{\overline{C}_p}Q_w = -\overline{\rho u_1}(T_w - T_{b}) - \frac{1}{\overline{C}_p}\frac{1}{RePr} \left.  \left( \lambda \frac{\partial T}{\partial x_1} \right) \right |_{x_{1,b}} + \int_{x_{1,b}}^0 q_l \, \mathrm{d} x_1.
\label{intTwall}
\end{equation}

Here $\overline{\rho u_1}$ is the mean $\rho u_1$ along the integral path.

The wall heat flux $Q_w$ is a function of various flowing quantities. Denote $\left<Q_w \right>|_{x_{1,b},T_{b}}$ as the mean heat flux conditional on $x_{1,b}$ and $T_{b}$ and $\left<Q_w \right>|_{x_{1,b}}$ as the mean heat flux conditional on $x_{1,b}$. It then implies 
\begin{equation}
\left<Q_w \right>|_{x_{1,b}}=\int\left<Q_w \right>|_{x_{1,b},T_{b}}pdf(T_{b})dT_{b},
\end{equation}
where $pdf(T_{b})$ is the PDF of $T_{b}$. Statistically we are interested in $\left<Q_w \right>|_{x_{1,b}}$. The temporal integral terms in Eq.~\eqref{headOnBc} and~\eqref{intTwall} are dominated by turbulent fluctuations. In the sense of statistical average, they are assumed to be negligible from modeling considerations. Consequently, $\left<Q_w \right>|_{x_{1,b},T_{b}}$ satisfies 

\begin{equation}
    \left< Q_w\right>|_{x_{1,b},T_{b}} = -{\overline{C}_p}(T_w - T_{b}) \left<\overline{\rho u_1}\right>|_{x_{1,b},T_{b}}  - \frac{1}{RePr} \left<  \lambda \frac{\mathrm{d} T}{\mathrm{d} x_1}  \right> |_{x_{1,b},T_{b}} + {\overline{C}_p}\left< \int_{x_{1,b}}^0 q_l \, \mathrm{d} x_1\right>|_{x_{1,b},T_{b}}.
\label{meanQwall}
\end{equation}

The solution of $\left<Q_w \right>|_{x_{1,b},T_{b}}$ from Eq.~\eqref{meanQwall} can equally be obtained from the following equation and boundary condition:
\begin{equation}
    \rho U_1 \frac{\mathrm{d} T}{\mathrm{d} x_1} = \frac{1}{C_p}\frac{1}{RePr}  \frac{\mathrm{d} \left( \lambda \frac{\mathrm{d} T}  {\mathrm{d} x_1} \right) } {\mathrm{d} x_1} + q_l, \quad \text{for } x_1 \geq x_{1,b},
\label{mean1dT}
\end{equation}

\begin{equation}
     \dot{m}_T T_{b} = -\frac{1}{\overline{C}_p}\frac{1}{RePr} \left.  \left( \lambda \frac{\partial T}{\partial \textit{x}_1} \frac{1}{\cos{\theta}} \right) \right |_{x_{1,b}} + \dot{m}_c c_{b},
\label{headOnBcNew}
\end{equation}
where the velocity $U_1$ is interpreted as the conditional mean of the turbulent velocity $u_1$.

\subsubsection{Velocity field}
To better understand the applicability of the present analyses, we decompose the upstream velocity into a wall tangential part and a wall normal part. Contribution from the tangential part is flow configuration dependent. Considering the wall normal velocity $\textbf{\textit{U}}$, in the flame upstream region ($x_1 \leq x_{1,u}$) it can be treated as a potential flow velocity $\textbf{\textit{U}}_p$, with the $x_1$ component $U_{1,p}$ and the radial component $U_{r,p}$. 

Physically $U_{1,p}$ equals to the mean $S_{\mathrm{zone}}$, i.e., $\left < S_{\mathrm{zone}}\right >$, at $x_1=x_{1,b}$. Thus $U_{1,p}$ and $U_{r,p}$ are
\begin{equation}
U_{1,p}\left(x_1\right)=-2 {\varepsilon}[{x}_1-\left({x}_{1,b}+\frac{\left <S_{\mathrm{zone}} \right >}{2 {\varepsilon}}\right)], \quad U_{r,p}\left({x}_r\right)={\varepsilon}{x}_r,
\label{potentialFlow}
\end{equation}
where ${\varepsilon}$ is the strain rate (normalized with $S_L^0/L$). As elaborated in Ref.~\citep{2018Analysis}, the mean flame zone speed $\left< S_{\mathrm{zone}} \right>$ can be modeled as
\begin{equation}
\left< S_{\mathrm{zone}} \right>=c_1+c_2 S_L(T_b).
\label{zone-speed}\end{equation}
$S_L$ in the above equation is the freely propagating flame speed at burnt temperature $T_b$. The parameters $c_1$ and $c_2$ satisfy $c_1+c_2=1$, implying that the movement of the near-wall flame is under the joint control of free propagation and wall confinement. Since the quantity $\left< S_{\mathrm{zone}} \right>$ describes the near-wall flame movement without much strong influence from the far field, $\left< S_{\mathrm{zone}} \right>$ is expected to be primarily determined by chemical kinetics, but not the flow configuration.

In the region between the flame boundary and wall ($x_1 \geq x_{1,b}$), $\textbf{\textit{U}}$ can be treated as the Homann flow~\citep{homann1936einfluss} with variable density. Specifically, $U_r$ is linearly proportional to $x_r$ and $U_1$ is $x_r$ independent. Therefore pressure $P$ satisfies $\cfrac{\partial}{\partial x_r}(\cfrac{\partial P}{\partial x_1}) = 0$, i.e., the pressure gradient $\partial P /\partial x_r$ is a constant along $x_1$. The constant $\partial P /\partial x_r$ can be further determined as $-\rho_u \varepsilon^2 x_r$ through the potential flow condition. The pressure $P$ above and the dynamic viscosity $\mu$ in the following equation is non-dimensionalized with $\rho_u  {S_L^0}^2$ and $\rho_u S_L^0 L$, respectively.

The momentum equation along $x_r$ is
\begin{equation}  
{\rho} {U}_1 \frac{\partial {U}_r}{\partial {x}_1} + {\rho} {U}_r \frac{\partial {U}_r}{\partial {x}_r} =
\rho_u \varepsilon^2 x_r + \frac{\partial}{\partial x_1} \left({\mu} \frac{\partial {U}_r}{\partial {x}_1}\right) +  \frac{\partial}{\partial {x}_r}\left({\mu}\frac{\partial {U}_r}{\partial {x}_r}\right) + \frac{{\mu}}{x_r}\frac{\partial U_r}{\partial x_r} - {\mu}\frac{U_r}{x_r^2}.
\label{momenteq}
\end{equation}
For the variable density case, we introduce a stretched coordinate ${\eta}$ after the Levy-Lees transformation~\citep{lees1956laminar} as
\begin{equation}
{\eta}=\left(\frac{2 {\rho}_u {\varepsilon}}{{\mu}_u}\right)^{\frac{1}{2}} \int_0^{{x}_{1}}\left(\frac{{\rho}}{{\rho}_u}\right) \mathrm{d} \zeta={\eta}({x}_1).
\end{equation}
Thus the stream function $\psi$, defined as ${\rho} {U}_1(x_1)=-\cfrac{1}{{x}_r} \cfrac{\partial \psi}{\partial {x}_r}$ and $\rho{U}_r(x_1,x_r)=\cfrac{1}{{x}_r} \cfrac{\partial \psi}{\partial {x}_1}$, can be written as 
\begin{equation} 
\psi = \sqrt{{\rho}_u{\mu}_u{\varepsilon} /2} x_r^2 f(\eta).
\end{equation}

In consistency with the assumption in nonreactive near-wall heat transfer analyses~\citep{lees1956laminar,libby1968stagnation}, $\rho \mu$ is assumed as constant, or the unity $({\rho} {\mu})/({\rho}_u {\mu}_u)$, because qualitatively the dependencies of viscosity and density on the temperature are opposite. Then, the momentum equation~\eqref{momenteq} becomes
\begin{equation}
{f}^{\prime \prime \prime} + {f} {f}^{\prime \prime}+\frac{1}{2}\left(\frac{{\rho}_u}{{\rho}}-{f}^{\prime 2}\right)=0.
\label{veleq1}\end{equation}

Under the isobaric condition $\rho = 1/(1 + \tau {T})$, where the heat release parameter $\tau = \left(T_{ad} - T_u\right)/T_{ad}$, Eq.~\eqref{veleq1} can be simplified as
\begin{equation}
{f}^{\prime \prime \prime} + {f} {f}^{\prime \prime}+\frac{1}{2}\left(1+\tau {T}-{f}^{\prime 2}\right) =0,
\label{veleq}
\end{equation}
together with the following boundary conditions
\begin{equation}
\begin{aligned}
 f(x_{1,b}) &= -\rho_b \left< S_{\mathrm{zone}} \right> \cos\theta /\left(2 {\rho}_u {\mu}_u {\varepsilon}\right)^{\frac{1}{2}},
\\ f(0) &=0 ,\\
 f^{\prime}(0) &=0.
\end{aligned}
\label{velbc}
\end{equation}
The velocity field from the function $f$ is coupled with the temperature equation ~\eqref{mean1dT}. Under the given boundary conditions, including the temperature boundary Eq.~\eqref{headOnBcNew}, the velocity boundary Eq.~\eqref{velbc}, and a prescribed wall temperature, the governing equations for $f$ and $T$ can be solved numerically. Because of nonlinear coupling, the equation set needs to be iterated until the variations of $f$ and $T$ are below a convergence criterion. The wall heat flux can then be quantified.

\subsection{Model validation}
\label{S5}

To validate the wall heat flux model in a general sense, we will validate the model in the laminar stagnation flame and the above two turbulent FWI configurations with different large-scale incoming flow orientations in this section.

Although the model is developed for turbulent flames, it also remains applicable under laminar conditions. Therefore, the proposed wall heat flux model is first validated using a quasi-one-dimensional laminar flame. To obtain the model solution of the wall heat flux, there are three input parameters, namely $\dot{m}_T$, $\dot{m}_c$, and $\delta_b$, i.e., the distance between $n_{b}$ and the wall. Also, as aforementioned, in the near-wall region, the flame zone speed $\left< S_{\mathrm{zone}} \right>$ is primarily determined by the chemical kinetics, or fuel-dependent, but flow configuration independent. The relation between $\left< S_{\mathrm{zone}} \right>$ and temperature can be predetermined.

\begin{figure}
\centering
     \begin{subfigure}[]{0.49\textwidth}
         \centering
         \includegraphics[width=\textwidth]{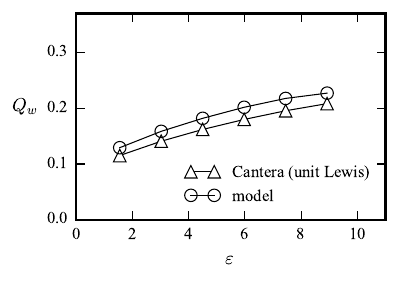}\vskip -8pt
         \label{fig:case1}
         \caption{}
     \end{subfigure}
     \begin{subfigure}[]{0.49\textwidth}
         \centering
         \includegraphics[width=\textwidth]{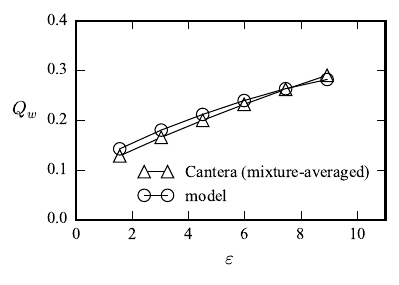}\vskip -8pt
         \label{fig:case2}
         \caption{}
     \end{subfigure}
\caption{Comparison of the modeled wall heat flux at various non-dimensional inlet strain rates $\varepsilon$ with the Cantera calculation: (a) results from the unity Lewis number assumption; (b) results from the mixture-averaged species transport model.}
   \label{fig:Laminar-com}
\end{figure}

To assess the model robustness, we examine the results from both the unit Lewis number model in Fig.~\ref{fig:Laminar-com} (a) and the mixture-averaged transport model in Fig.~\ref{fig:Laminar-com} (b). It can be seen that the agreement is satisfactory across a broad range of the inlet strain rates, even for such detailed chemical kinetics and the complex transport model. Moreover, despite the non-zero near-wall heat release rate from the mixture-averaged transport model, the model prediction will not be clearly influenced because most of the heat is still released from the flame zone. 

\subsubsection{Turbulent wall normal flushing flame}

To justify the general applicability of the present wall heat flux model, we adopt the multi-step kinetics results in~\cite{zhao2022near} to validate the model. From each wall point, the local wall heat flux can be directly computed. Meanwhile, moving along the wall normal direction, the closest flame patch is identified as the corresponding flame patch. 

\begin{figure}
    \centering
    \includegraphics[width=0.8\textwidth]{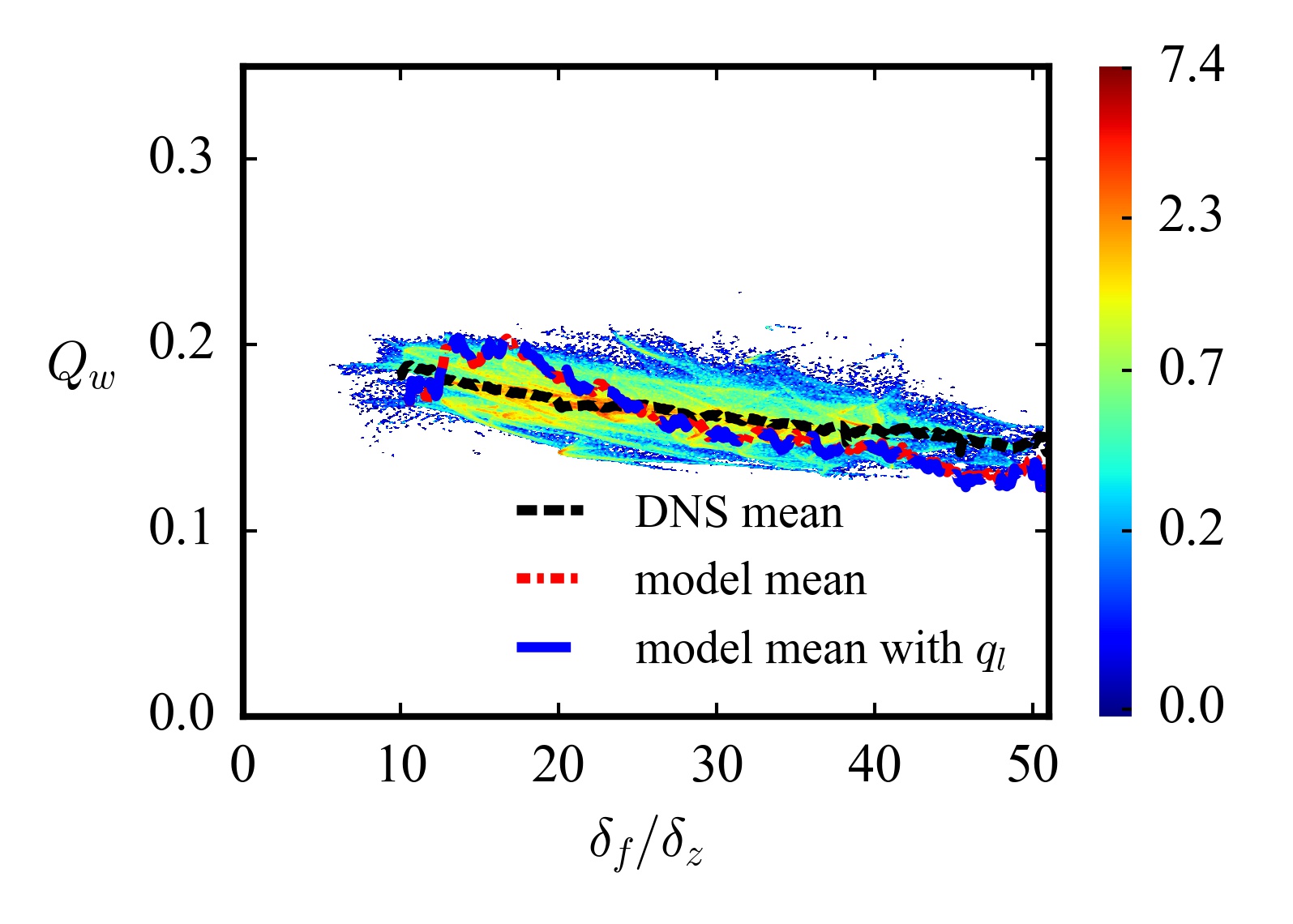}
    \caption{Model predicted wall heat flux with respect to the flame-wall distance, both with $q_l$ correction and without $q_l$, in comparison with the DNS joint PDF and its conditional mean.}
    \label{fig:DXres}
\end{figure}

To evaluate the wall heat flux from the model, the mean flame zone speed $\left< S_{\mathrm{zone}} \right>$ parameters $c_1$ and $c_2$ in Eq.~\eqref{zone-speed} for head-on flames are set as $c_1=0.2$ and $c_2=0.8$ from DNS database fitting. From the previous discussion of 
$\left< S_{\mathrm{zone}} \right>$, the determined $c_1$ and $c_2$ values are generally valid for various hydrogen-air FWI configurations at atmospheric pressure. Since along the wall parallel direction the influence from the flame is much weaker than the wall normal part, in the model implementation, the lateral heat transfer can be locally evaluated as the nonreactive case, for instance, by the turbulent heat conductivity and temperature gradient. In the following, model results both with $q_l$ and without $q_l$ will be compared. Fig.~\ref{fig:DXres} shows the model-predicted mean wall heat flux with respect to the flame-wall distance $\delta_{f}$, together with the DNS joint PDF and its conditional mean. It can be seen that the DNS result and model predictions with $q_l$ correction and without $q_l$ almost collapse, justifying the model performance in this flushing flame configuration. Consistent with the small mean value of wall parallel heat transfer in Fig.~\ref{fig:ql} (a), the influence of $q_l$ on the wall heat flux is insignificant.

\subsubsection{Turbulent inclined sweeping flame}
\label{swq}
In this inclined sweeping flame quenching scenario, the flame zone is not aligned with the wall boundary. According to Eq.~\eqref{headOnBcNew}, to obtain the model solution, the orientation of the local flame elements has to be considered. For the present inclined sweeping FWI case, the elevated ambient pressure leads to the change of chemical reaction rates, which then will slightly change the fitting parameters of the mean flame zone speed $\left< S_{\mathrm{zone}} \right>$ to $c_1=0.3$ and $c_2=0.7$.

\begin{figure}
\centering
     \begin{subfigure}[]{0.51\textwidth}
         \centering
         \includegraphics[width=\textwidth]{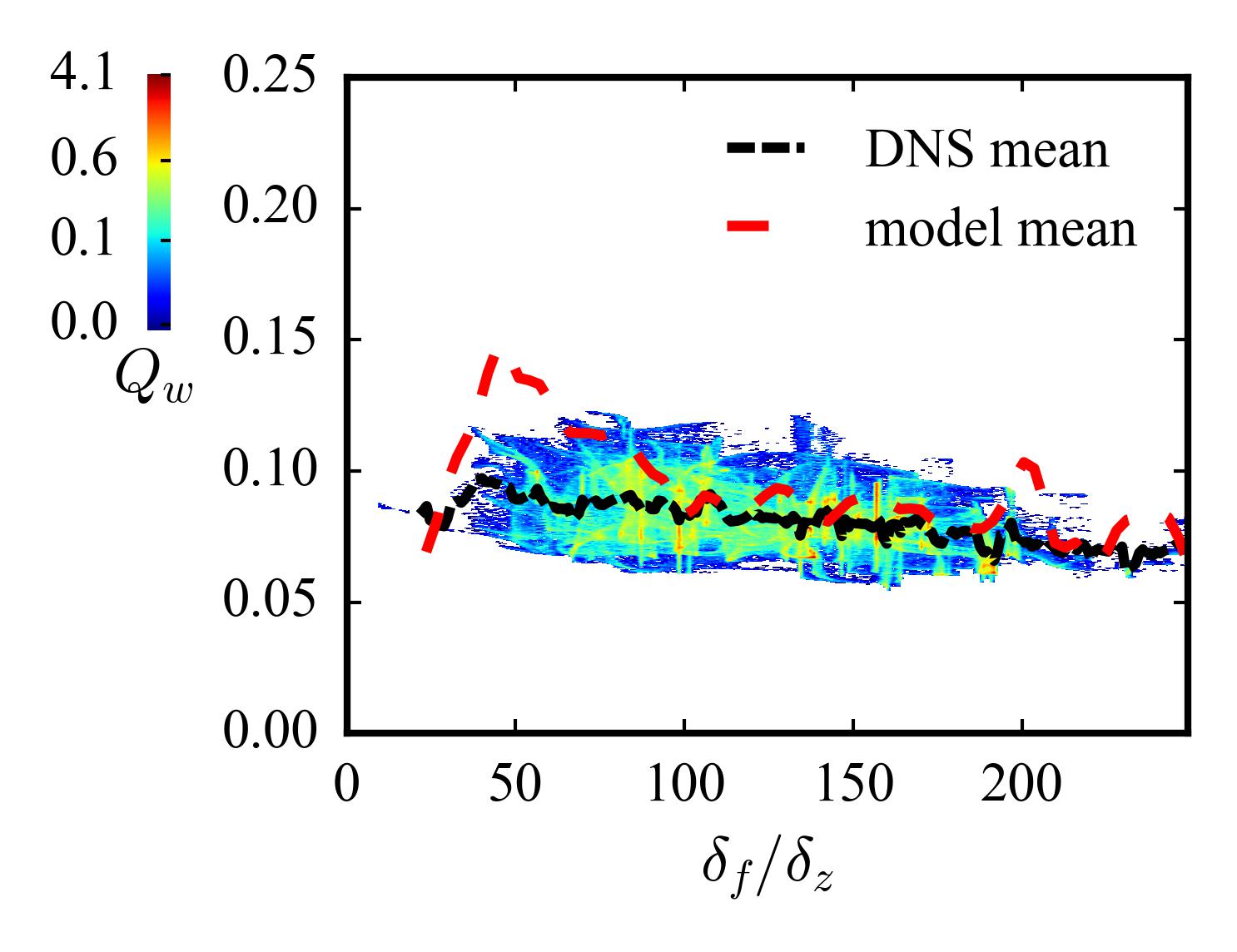}\vskip -8pt
         \caption{}
     \end{subfigure}
     \begin{subfigure}[]{0.46\textwidth}
         \centering
         \includegraphics[width=\textwidth]{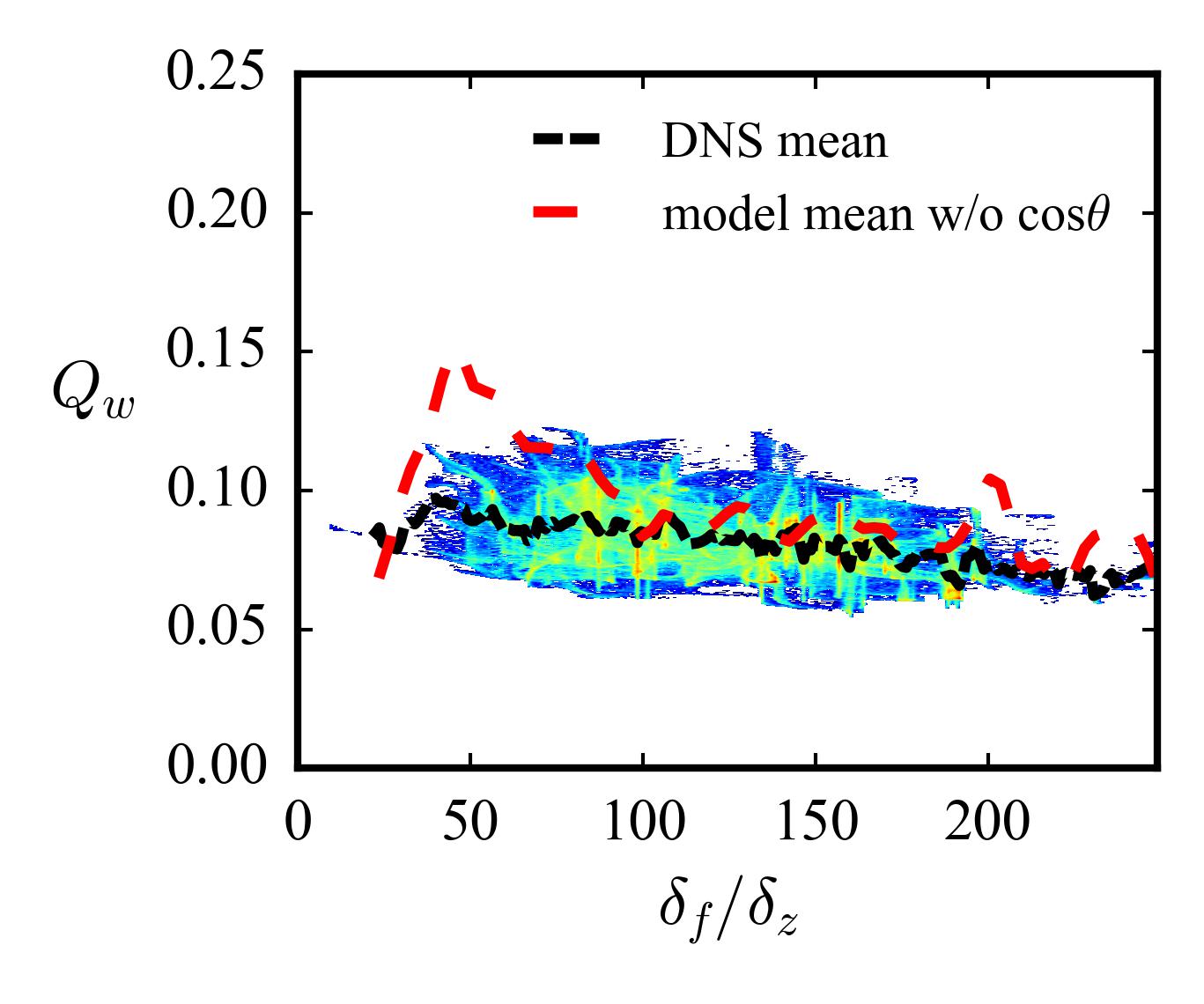}\vskip -8pt
         \caption{}
     \end{subfigure}
     \begin{subfigure}[]{0.46\textwidth}
         \centering
         \includegraphics[width=\textwidth]{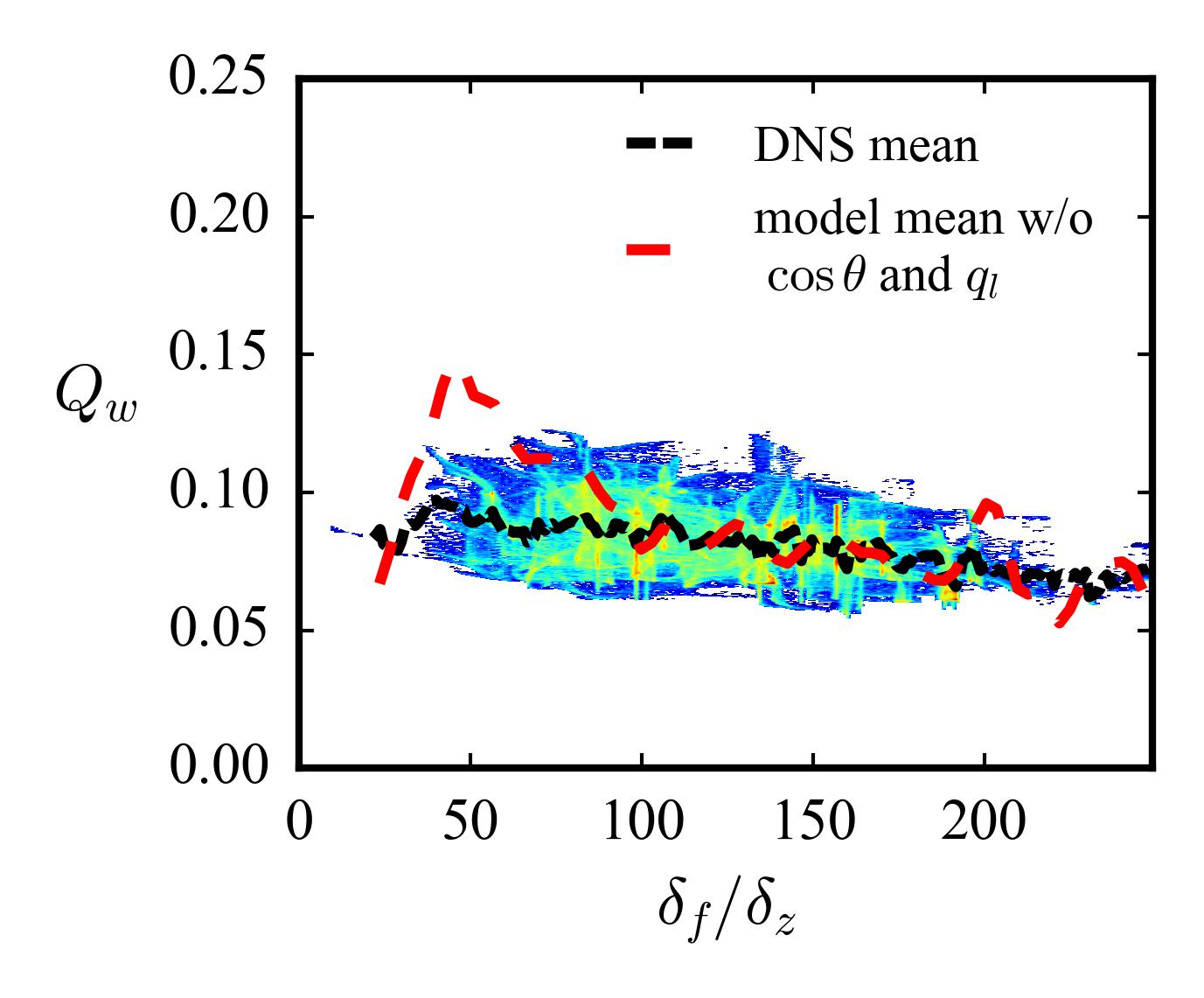}\vskip -8pt
         \caption{}
     \end{subfigure}
     \begin{subfigure}[]{0.46\textwidth}
         \centering
         \includegraphics[width=\textwidth]{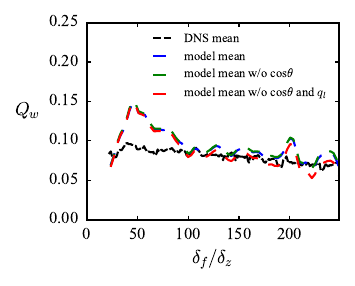}\vskip -8pt
         \caption{}
     \end{subfigure}
\caption{Wall heat flux predictions for the inclined sweeping FWI case, in comparison with the DNS joint PDF and its conditional mean from: (a) complete model solution; (b) model solution without the influence from the local flame element orientation $\cos\theta$; (c) model solution without influences from flame element orientation $\cos\theta$ and the wall parallel heat transfer $q_l$; (d) conditional means of the DNS and model solutions with different simplifications.}
\label{fig:SWQ}
\end{figure}

Fig.~\ref{fig:SWQ}~(a) shows the model predictions compared with DNS results. When the flame is close to the wall ($\delta_f/\delta_z <100$), the modeled wall heat flux is a bit overpredicted, while when the flame is away from the wall, the agreement is satisfactory. It is also interesting to test the model robustness with possible simplifications. Fig.~\ref{fig:SWQ}~(b) presents the model result without correction from the flame element orientation, i.e., set $\cos{\theta} = 1$ in Eq.~\eqref{headOnBcNew} and~\eqref{velbc}. It can be seen that the result is almost invariant, implying the weak influence of the flame element orientation, since the dominant part of the wall heat flux aligns with the wall normal. Furthermore, if the wall parallel heat transfer $q_l$ in Eq.~\eqref{mean1dT} is also excluded, the model result is shown in Fig.~\ref{fig:SWQ}~(c). The difference is still small, and only a slight deviation exists at the far end of the solution curve. Fig.~\ref{fig:SWQ}~(d) is provided for comparison of the different simplifications of the model. In summary, the model is capable of capturing the dominant part of the wall heat flux with good robustness, even without correction from the local flame element orientation and wall parallel heat transfer.

\section{Conclusions}
The present study focuses on the fundamental physics of the premixed near-wall flame, together with wall heat flux modeling for the practical application purpose. Results from two different flame configurations are compared, including the wall normal flushing flame and inclined sweeping flame. 

First, the laminar stagnation flame is considered as a reference. It is found that the species enthalpy diffusion term is insignificant compared to the temperature diffusion term. A second-order tensor induced by the skin friction vector, as a degenerated velocity gradient tensor on the wall, has been introduced to study the interaction between the flame and velocity on the wall. After decomposing into a symmetric and antisymmetric part, the wall vorticity, flame normal and tangential strain rates can be accordingly defined. It is found that the large curvature magnitudes typically correspond to small vorticity magnitudes, which are linked with a wall vortex pair structure. Statistically, oxygen-based progress variable contour lines exhibit larger curvatures than the hydrogen-based progress variable contour lines. A negative correlation is also observed between the flame curvature and tangential strain rate. On the wall boundary the alignment between the progress variable gradient and the most compressive strain direction conditional on large gradient magnitude is strong, but weakens as the gradient decreases. To characterize the near-wall flame structure, a newly proposed species alignment index suggests that when approaching the wall, isosurfaces of different species tend to be more misaligned. Moreover, in both configurations, near-wall heat transfer is primarily governed by the wall-normal component, while the wall-parallel part is unimportant.

Based on the above analyses, we proposed a wall heat flux model for practical application purposes under realistic conditions, including flame with finite thickness, complex chemical kinetics, non-negligible near-wall reactions, and variable flame orientation relative to the wall, etc. For all test cases, predictions of this new wall heat flux model agree well with DNS results. The model is also ideally robust because of the insignificant influences from the species transport model, wall parallel heat transfer, and local flame element orientation. The model inputs are expected to be resolution-independent or weakly dependent due to the integration calculations. In this sense, this model can promisingly be integrated into existing LES solvers.

\backsection[Funding]{The funding support from the National Natural Science Foundation of China (NSFC) under Grant No. 12272229 is acknowledged. }

\backsection[Declaration of interests]{The authors declare that they have no known competing financial interests or personal relationships that could have appeared to influence the work reported in this paper.
}

\backsection[Author contributions]{\textbf{KL}: formal analysis, calculation, validation, and wrote the paper.
\textbf{CG}: formal analysis and calculation.
\textbf{ZZ}: organizing DNS data, helpful discussion.
\textbf{HW}: reviewed and revised the paper.
\textbf{LW}: project design, formal analysis, methodology, funding acquisition, reviewed and revised the paper.}

\bibliographystyle{jfm}
\bibliography{jfm}

\end{document}